\documentclass[11pt,a4paper]{article}

\usepackage[a4paper,margin=2.0cm]{geometry}

\usepackage[numbers,sort&compress]{natbib}
\usepackage{graphicx} 
\usepackage{quantikz}

\usepackage{amsmath,amssymb,amsfonts}
\usepackage{algorithmic}
\usepackage{graphicx}
\usepackage{hyperref}
\usepackage{subcaption}
\usepackage{textcomp}
\usepackage{xcolor}
\usepackage{placeins}
\usepackage{float}
\usepackage{dsfont}
\usepackage{subcaption}
\usepackage{rotating}
\usepackage{booktabs}
\usepackage{tikz}
\usetikzlibrary{
  arrows.meta,
  positioning,
  shapes.multipart,
  shapes.geometric,
  fit,
  backgrounds,
  calc,
  decorations.pathreplacing,
  matrix
}

\tikzset{
  umlclass/.style={
    rectangle split, rectangle split parts=#1,
    draw=black!80, fill=blue!5,
    minimum width=5.2cm,
    font=\small,
    inner sep=6pt,
    rounded corners=2pt,
  },
  umlclass/.default=3,
  abstractclass/.style={
    umlclass=#1, fill=yellow!8, draw=orange!60!black,
  },
  abstractclass/.default=3,
  inherits/.style={
    -{Triangle[open, length=8pt, width=6pt]}, thick,
  },
  uses/.style={
    -{Stealth[length=6pt]}, dashed, thick,
  },
  composes/.style={
    -{Diamond[length=8pt, width=6pt, fill=black]}, thick,
  },
  actor/.style={
    rectangle, draw, fill=blue!10, minimum width=2.2cm, minimum height=0.8cm,
    font=\small\bfseries, rounded corners=3pt,
  },
  seqarrow/.style={
    -{Stealth[length=5pt]}, thick,
  },
  seqreturn/.style={
    -{Stealth[length=5pt]}, thick, dashed,
  },
  pipebox/.style={
    rectangle, draw=black!70, fill=#1, minimum width=3.8cm,
    minimum height=1.0cm, font=\small, align=center, rounded corners=3pt,
  },
  pipebox/.default={blue!10},
  encbox/.style={
    rectangle, draw=black!60, fill=gray!8, minimum width=1.1cm,
    minimum height=0.9cm, font=\small\ttfamily, align=center,
  },
}

\title{CO$\Phi$ -- COmpiling PHotonics into dIgital quantum circuits}
\author{
Grzegorz Czelusta$^1$\thanks{grzegorz.czelusta@uj.edu.pl}, 
Marcin Markiewicz$^{2,3,4}$\thanks{mmarkiewicz@kplabs.pl},  
Jakub Mielczarek$^1$\thanks{jakub.mielczarek@uj.edu.pl}, 
Artur Miroszewski$^{1,5}$\thanks{artur.miroszewski@esa.int} \\ \\
$^{1}$Institute of Theoretical Physics, Jagiellonian University, \\
{\L}ojasiewicza 11, 30-348 Cracow, Poland\\
$^{2}$ Institute of Theoretical and Applied Informatics, Polish Academy of Sciences,\\ 
ul. Ba{\l}tycka 5, 44-100 Gliwice, Poland\\
$^{3}$ International Centre for Theory of Quantum Technologies \\
(ICTQT), University of Gda\'nsk, 80-308 Gda\'nsk, Poland \\
$^{4}$ KP Labs, Bojkowska 37J, 44-100 Gliwice, Poland \\
$^{5}$ $\Phi$-lab, European Space Agency (ESA/ESRIN), Frascati, Italy
}

\date{\today}

\begin{document}

\maketitle

\begin{abstract}

Photonic and qubit-based (digital) approaches to quantum computing are 
largely regarded as separate paradigms. This article aims to bridge these 
paradigms by developing a method for mapping quantum-optical systems 
onto qubit-based quantum circuits. 

The key ingredient enabling this mapping is the introduction of a truncation 
of the Fock space describing single-mode optical states. Within this framework, 
representations of basic photonic operations are developed, including the 
phase-rotation operator, the displacement operator, the squeezing operator, 
and the beam splitter. Explicit quantum circuit constructions are provided 
for Fock-space truncations of dimension $N=2$ and $N=4$. We discuss benchmarking 
of the implemented operations and identify the regime of applicability 
and the limitations of the proposed method.

The presented results can be used to simulate quantum optical circuits on 
qubit-based quantum computers, with special relevance to Gaussian Boson Sampling in the low-photon-number regime.
To demonstrate the practical applicability of this approach, we complement the 
theoretical framework with with a Python-based quantum programming tool CO$\Phi$ 
that compiles Blackbird quantum-optics programs into qubit-based OpenQASM circuits 
using a truncated Fock-space encoding. The compiler is available on GitHub at \url{https://github.com/Quantum-Cosmos-Lab/cophi}.
\end{abstract}

\section{Introduction}

In quantum optics, one naturally works with an infinite-dimensional Hilbert space,
a feature that plays a central role in quantum information processing. In particular,
the availability of an unbounded set of photonic excitation levels enables
continuous-variable models of quantum computation. 

Specifically, states of single-mode light are described within an infinite-dimensional
Hilbert space known as the \emph{Fock space}. Its basis states are labeled by the photon
occupation number $n$ and denoted by $\ket{n}$, such that the Hilbert space reads:
\begin{equation}
\mathcal{H} = \mathrm{span}\{\ket{0}, \ket{1}, \ket{2}, \ldots\}.
\end{equation}

This infinite-dimensional structure underlies a wide range of quantum optical
protocols, including sampling \cite{Aaronson:2010kxa} as well as quadrature-based 
quantum communication \cite{Grosshans2002, Weedbrook2012, Braunstein2005} and sensing schemes  \cite{Caves1981, Pirandola2018, LIGO2013}. 
A prominent example is Gaussian Boson Sampling (GBS), for which the possibility of 
achieving quantum computational advantage has been rigorously demonstrated~\cite{Hamilton:2017cbx,Brod2019GBSComplexity, Kruse2019}. Large-scale photonic realizations have subsequently demonstrated the experimental reach of GBS architectures \cite{Zhong2020,Madsen2022,Deshpande2022}. The classical simulation complexity of GBS has also motivated specialized algorithms and simulation methods \cite{Quesada2022}.

In contrast, currently available universal quantum computers are operating on 
finite-dimensional quantum registers composed of qubits. An $m$-qubit register
is described by a Hilbert space of dimension $2^m$, which fundamentally limits
the direct representation of photonic degrees of freedom. Establishing a faithful
connection between these two paradigms, therefore, requires a systematic mapping
between Fock states and qubit registers.

In this article, we investigate the feasibility of implementing quantum optical
circuits using a qubit-based quantum register, with the goal of linking the 
two paradigms. Such an approach necessarily requires embedding photonic processes, 
which are intrinsically infinite-dimensional, into a finite-dimensional Hilbert 
space accessible to qubit-based quantum computers. This, in turn, calls for the 
introduction of a well-controlled truncation of the Fock space.

Here, we adopt a strategy that preserves unitarity at every stage of the
construction. This is achieved by working entirely within a truncated Fock space
equipped with appropriately defined creation and annihilation operators, 
which converges with the approach developed in Refs. \cite{Trzetrzelewski:2004nz,Trzetrzelewski:2003sz}. 

Furthermore, we show that the finite-dimensional truncated Hilbert space admits a natural algebraic interpretation in terms of \(q\)-deformed oscillator algebras \cite{ArikCoon1976, Biedenharn1989, Macfarlane1989, KulishDamaskinsky1990}. The concept of \(q\)-deformation has been widely explored in contexts ranging from fractional quantum statistics to candidate quantum gravity effects \cite{ Greenberg1990, MajorSmolin1996, DupuisGirelli2013}, and provides a useful algebraic framework for introducing finite-dimensional representations and occupation-number cutoffs. When the deformation parameter is chosen to be a root of unity, the corresponding representation becomes finite-dimensional, while the standard bosonic algebra is recovered in the limit \(q\to1\).

Within this framework, we not only establish a mapping between photonic states and
qubit registers but also explicitly represent the action of fundamental quantum
optical elements, including phase rotations, displacement operations, squeezing,
and beam splitter, in terms of qubit-based quantum circuits. As a result,
both quantum states of light and their associated optical transformations admit an
explicit circuit-level description. 

In general, mappings of bosonic or finite-dimensional degrees of freedom onto qubit registers have been studied in the context of digital quantum simulation, including oscillator, vibrational, and general \(d\)-level systems \cite{Somma2002,Somma2003,Macridin2018PRA,Sawaya2020,DiMatteo2021}. Related developments include digital approaches to molecular vibronic problems \cite{SawayaHuh2019}, bosonic software frameworks \cite{Stavenger2022}, fermion-to-qubit representations of linear optics \cite{Chin2024}, and recent gate-based emulations of boson sampling \cite{Zalone2026}. A complementary block-encoding approach for bosonic ladder operators has recently been developed in Ref. \cite{Simon:2025ladderoperatorblock}. A distinct route avoids a purely qubit-based encoding and instead employs native hybrid oscillator–qubit processors \cite{Liu2026hybrid}.
Furthermore, an alternative approach to emulate 
optical circuits on digital quantum register has recently been developed \cite{Mohan:2024gvv,Brasil:2025afa}. In the approach the Gray code mapping 
of the Fock states to the qubit states has been applied. This allows 
decomposition of the bosonic creation and annihilation operators into the 
Pauli strings. Then quantum circuits for the exponentiation of the individual 
Pauli strings have been constructed. This approach allows simulating  one 
of the most characteristic bosonic effects \cite{Mohan:2024gvv}, which is the Hong-Ou-Mandel 
interference \cite{HOM1987}. Furthermore, the squeezed states in the 
framework have been analyzed in Ref. \cite{Li:2025qjz}.

Importantly, here we not only provide an alternative and optimized truncation method but supplement our method with the Python-based quantum 
compiler CO$\Phi$. The compiler enables translating the Blackbird quantum-optics 
programs  \cite{ Killoran2019, Bromley2020} into qubit-based OpenQASM circuits \cite{Cross2017, Cross2022}.
The compiler has been made available on GitHub repository:
\url{https://github.com/Quantum-Cosmos-Lab/cophi}.

The remainder of this article is organized as follows.
In Sec.~II, we review the basic elements of quantum optics relevant to our work,
focusing on the description of single-mode radiation in Fock space and on the
fundamental optical operators that form the building blocks of quantum optical
circuits. In Sec.~III, we introduce the truncated Fock space formalism and define
the corresponding creation and annihilation operators, discussing their algebraic
properties and deviations from the canonical bosonic commutation relations.

In Sec.~IV, we establish the connection between the truncated Fock space and
$q$-deformed oscillator algebras, showing how a finite-dimensional Hilbert space
arises naturally from the deformation parameter and how the standard bosonic
algebra is recovered in the appropriate limit. In Sec.~V, we present the encoding
scheme that maps photonic Fock states onto qubit-based quantum registers, both for
single-mode and multi-mode systems.

Sections~VI and VII are devoted to explicit constructions of quantum circuits
implementing truncated quantum optical operations. In Sec.~VI, we analyze the
case of the lowest-dimensional truncation $N=2$, corresponding to a single-qubit
representation, and derive circuit decompositions for phase shifts, displacement
operations, squeezing, and beam splitters. In Sec.~VII, we extend this analysis to
the higher-dimensional truncation $N=4$, corresponding to a two-qubit register,
and provide explicit gate-level implementations of the same set of optical
operations.

In Section~VIII, we benchmark the implemented operations and identify the range 
of applicability and limitations of the proposed method. Section~IX presents the 
implementation of the method in the CO$\Phi$ compiler, outlines the code structure, 
and provides practical information for users.

Finally, in Section X, we summarize our results and discuss their implications for the
simulation of quantum optical circuits, including truncated Gaussian Boson
Sampling architectures, on universal qubit-based quantum computers.

The article is supplemented by two appendices. Appendix A reviews the basics of 
the $q$-deformed oscillator, while Appendix B summarises the definitions of the 
quantum gates used throughout the article.

In this article, we use units in which $\hbar = 1$.

\section{Standard quantum optics}

In this section, we briefly summarize the essential concepts of quantum 
optics that are relevant for the present work. We focus on the quantum 
description of a single-mode light, whose states are defined in an infinite-dimensional 
Hilbert space known as the Fock space. Within this framework, optical 
transformations are described by operators acting on the photon-number 
states.

We introduce the basic operators that constitute the building blocks of quantum 
optical circuits and form the subject of our analysis. These include the phase 
rotation operator, the displacement operator, the squeezing operator, and the 
symmetric beam splitter. Their properties and algebraic structure are reviewed 
in the context of single-mode and two-mode quantum optics, providing the foundation 
for the constructions developed in the subsequent sections.

\subsection{Single mode of light}

The single mode of light is described by the harmonic oscillator, 
for which the Hamiltonian is: 
\begin{equation}
\hat{H}= \omega \left( \hat{a}^{\dagger}\hat{a}+\frac{1}{2}\hat{\mathbb{I}}\right),
\end{equation}
where $\omega$ denotes angular frequency 
and $\hat{a}^{\dagger}$ and $\hat{a}$ are creation and annihilation 
operators respectively, satisfying the standard commutation relations:
\begin{align}
\label{CCRcom}
[\hat{a}, \hat{a}^{\dagger}] &= \hat{\mathbb{I}}.
\end{align}

The operators allow the introduction of the 
occupation number operator $\hat{N} :=\hat{a}^{\dagger}\hat{a}$, 
which satisfies the following eigenequation:
\begin{equation}
\hat{N}\ket{n}=n\ket{n},
\end{equation}
where $n \in \mathbb{N}$ is the occupation number. Action of the 
creation and annihilation operators on the occupation number eigenstates
is:
\begin{align}
\hat{a}^{\dagger}\ket{n} &= \sqrt{n+1}\ket{n+1}, \\
\hat{a}_i\ket{n} &= \sqrt{n}\ket{n-1}.
\end{align}

\subsection{Phase shift}

The phase shift operator is defined as the following exponentiation of the 
occupation number operator $\hat{N}$:
\begin{equation}
\hat{U}(\theta)= \exp \left(-i\theta \hat{N} \right),
\label{PSOperator}
\end{equation}
so that, action on the eigenstates $\ket{n}$ is given by: 
\begin{equation}
\hat{U}(\theta) \ket{n}= \exp \left(-i\theta n\right)\ket{n}.
\end{equation}

Under the phase shift, the annihilation and creation operators, transform as follows:
\begin{align}
\hat{U}(\theta)^{\dagger}\hat{a}\hat{U}(\theta) &=\hat{a}e^{-i\theta}, \label{Ua} \\
\hat{U}(\theta)^{\dagger}\hat{a}^{\dagger}\hat{U}(\theta) &=\hat{a}^{\dagger}e^{i\theta}.
\label{Uadag}
\end{align}

\subsection{Glauber coherent states}

The Glauber coherent states are defined as the eigenstates 
of the annihilation operators:
\begin{align}
\hat{a} \ket{\alpha} = \alpha \ket{\alpha}, 
\end{align}
where $\alpha=r\,e^{i\varphi} \in \mathbb{C}$. The states 
can generated from vacuum with use of the displacement operator, 
which is defined as follows:
\begin{equation}
\hat{D}(\alpha) := \exp \left(\alpha \hat{a}^{\dagger}-\alpha^* \hat{a}\right),
\end{equation}
so that :
\begin{align}
\ket{\alpha} = \hat{D}(\alpha) \ket{0} = e^{-|\alpha|^2/2} \sum_{n=0}^{\infty} \frac{\alpha^n}{\sqrt{n!}}\ket{n}. 
\end{align}

The properties of the displacement operator include:
\begin{align}
\hat{D}^{\dagger}(\alpha) &= \hat{D}(-\alpha), \\
\hat{D}^{\dagger}(\alpha)\hat{a} \hat{D}(\alpha) &= \hat{a}+\hat{\mathbb{I}}\alpha, \\
\hat{D}(\alpha)\hat{D}(\beta) &=e^{(\alpha\beta^*-\alpha^*\beta)/2} \hat{D}(\alpha+\beta).
\end{align}

For the further utility, it is worth observing that:
\begin{equation}
\hat{U}(\theta)^{\dagger} \left(\alpha \hat{a}^{\dagger}-\alpha^* \hat{a}\right) 
\hat{U}(\theta) = \alpha e^{i\theta}  \hat{a}^{\dagger}- \left(\alpha e^{i\theta} \right)^* \hat{a},
\end{equation}
where Eq. \ref{Ua} and Eq. \ref{Uadag} have been used. From here, 
it is straightforward that:
\begin{equation}
\hat{U}(\theta)^{\dagger} \hat{D}(\alpha)
\hat{U}(\theta) = \hat{D}(e^{i\theta}\alpha).
\end{equation}
From here, by fixing $\theta=-\varphi$, we find that:  
\begin{equation}
\hat{U}(-\varphi)^{\dagger} \hat{D}(\alpha)
\hat{U}(-\varphi) = \hat{D}(|\alpha|),
\end{equation}
so that the displacement operator for an arbitrary $\alpha$ can be 
decomposed in terms of the displacement operator with real $|\alpha|$
and the phase shifts which encode the phase information:
\begin{equation}
\hat{D}(\alpha) = \hat{U}(\varphi)^{\dagger} \hat{D} (|\alpha|)\hat{U}(\varphi).
\label{DDecomposition}
\end{equation}
This relation turns out to be useful in finding quantum circuit representation 
for the displacement operator.  

\subsection{Squeezed states}

The squeezing operator is defined as:
\begin{equation}
\hat{S}(z)= \exp \left[ (z^*\hat{a}^2-z (\hat{a}^{\dagger})^2)/2\right],
\end{equation}
where the squeezing parameter $z =re^{i\phi} \in \mathbb{C}$.
Here, $r \geq 0$ is the squeezing parameter and $\phi$ determines 
the squeezing direction in phase space.

Action of the operator $\hat{S}(z)$ onto the vacuum state, 
leads to the squeezed vacuum state, which, in the Fock basis, 
can be written as: 
\begin{equation} 
\ket{z} = \frac{1}{\sqrt{\cosh r}} \sum_{n=0}^{\infty} \frac{\sqrt{(2n)!}}{2^n n!} \left(-e^{i\phi}\tanh r\right)^n \ket{2n}. 
\end{equation}

\subsection{Beam splitter}

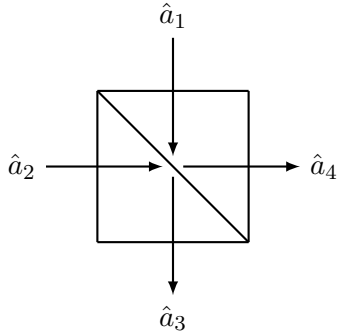
\begin{figure}[ht!]
\centering
\begin{tikzpicture}

\node[] (N0)  at (0cm,0cm) {};
\node[] (N1)  at (-1cm,1cm) {};
\node[] (N2)  at (1cm,1cm) {};
\node[] (N3)  at (1cm,-1cm) {};
\node[] (N4)  at (-1cm,-1cm) {};

\draw[thick] (N1.center) -- (N2.center);
\draw[thick] (N2.center) -- (N3.center);
\draw[thick] (N3.center) -- (N4.center);
\draw[thick] (N4.center) -- (N1.center);

\draw[thick] (N1.center) -- (N3.center);

\node[] (NL)  at (-2cm, 0cm) {$\hat{a}_2$};
\node[] (NR)  at (2cm, 0cm) {$\hat{a}_4$};
\node[] (NU)  at (0cm, 2cm) {$\hat{a}_1$};
\node[] (ND)  at (0cm, -2cm) {$\hat{a}_3$};

\draw[thick,-latex] (NL) -- (N0);
\draw[thick,-latex] (NU) -- (N0);
\draw[thick,latex-] (NR) -- (N0);
\draw[thick,latex-] (ND) -- (N0);

\end{tikzpicture}
\caption{Schematic diagram of the quantum beam splitter.}
\label{Fig:Splitter}
\end{figure}

A beam splitter (see Fig. \ref{Fig:Splitter}) is characterized by two 
input channels (denoted as $1,2$) and two output channels (denoted as $3,4$). 
Associated with these channels are the Fock spaces:
\begin{equation}
\mathcal{H}_i = \text{span}\bigl\{ \ket{0}_i, \ket{1}_i, \ket{2}_i, \ldots \bigr\},
\end{equation}
which describe the states of light.
In these spaces, we can define the creation operators $\hat{a}^{\dagger}_i$ and annihilation operators $\hat{a}_i$, for $i=1,2,3,4$, satisfying the standard commutation relations:
\begin{align}
\label{CCRcomms}
[\hat{a}_i, \hat{a}^{\dagger}_j] &= \hat{\mathbb{I}}\,\delta_{ij}, \\
[\hat{a}_i, \hat{a}_j] &= 0 = [\hat{a}^{\dagger}_i, \hat{a}^{\dagger}_j].
\end{align}
Moreover, the number operators are defined as:
\[
\hat{N}_i := \hat{a}^{\dagger}_i \hat{a}_i.
\]

The transformation operator for the symmetric 
beam splitter is given by:
\begin{equation}
\hat{U}_{BS}(\theta,\phi) =  \exp \left[ \theta\left(e^{i\phi}\hat{a}^{\dagger}_1\otimes
\hat{a}_2-e^{-i\phi}\hat{a}_1\otimes \hat{a}_2^{\dagger}    \right)   \right],
\label{BSOperator}
\end{equation}
which is parametrized by the two angles $\theta$ and $\phi$
The case of symmetric beam splitter is obtained for $\theta = \frac{\pi}{4}$
and $\phi = 0$.

The phase $\phi$ dependence can be extracted from the (\ref{BSOperator}) 
operator with the use of the phase shift operator (\ref{PSOperator}), 
leading to the following decomposition: 
\begin{equation}
\hat{U}_{BS}(\theta,\phi)=(\hat{U}(\phi)^\dagger \otimes \hat{\mathbb{I}})
\hat{U}_{BS}(\theta,0)(\hat{U}(\phi)\otimes\hat{\mathbb{I}}), 
\label{UBSDecomposition}
\end{equation}
which can be proven with the use of Eq. \ref{Ua} and Eq. \ref{Uadag}. 

From here, one can find the following transformation rule of the 
annihilation operators:
\begin{equation}
\begin{pmatrix} \hat{a}_3 \\ \hat{a}_4  \end{pmatrix}
=  \hat{U}_{BS}^{\dagger}\begin{pmatrix} \hat{a}_1 \\ \hat{a}_2  \end{pmatrix} \hat{U}_{BS} =  \begin{pmatrix} \cos\theta & e^{i \phi}\sin{\theta} \\ 
-e^{-i \phi}\sin \theta&  \cos\theta \end{pmatrix} \begin{pmatrix} \hat{a}_1 \\ \hat{a}_2  \end{pmatrix}.
\end{equation}

The above transformation is a simple example of a general passive linear optical transformation (representing an optical multiport), characterized by the fact that each single-mode operator transforms independently:
\begin{equation}
    \hat a_i^{\textrm{out}}=\sum_{j=1}^{n_{\textrm{modes}}}U_{ij}\hat a_j^{\textrm{in}},
\end{equation}
in which  $U_{ij}$ is a matrix representation of the multiport acting on $n_{\textrm{modes}}$.

It is worth mentioning that while the BS operation is performed, 
the input Hilbert space is no longer involved. Therefore, we 
can perform the following identification of the Hilbert spaces 
associated with the input and output states:
\begin{align}
\mathcal{H}_1 &\rightarrow \mathcal{H}_3, \\
\mathcal{H}_2 &\rightarrow \mathcal{H}_4, 
\end{align}
so that it is sufficient to operate on the $\mathcal{H}_1\otimes\mathcal{H}_2$
tensor product space. This is consistent with the deterministic 
evolution of the quantum state through the BS. 

\section{Cut-off Fock space}

The central idea of the finite qubit representation of the optical Fock space is to consider finite-dimensional 
truncation of the space \cite{PeggBarnett1989, Miranowicz2001}:
\begin{equation}
\mathcal{H} = \text{span}\{\ket{0}, \ket{1}, \ket{2}, \ldots, \ket{N-1}\}.
\end{equation}
Here, the truncation of the Fock space to $N$ dimensions is introduced by the 
cut-off annihilation and creation operators\footnote{In order to distinguish 
these operators from the case of standard infinite-dimensional case, we denote
them as $\hat{b}$ and $\hat{b}^{\dagger}$ instead of $\hat{a}$ and $\hat{a}^{\dagger}$, as introduced in Sec. 2.}, the matrix representation 
of which is the following: 
\begin{align}
\hat{b} &=
\begin{pmatrix}
0 & 1 & 0 & 0 & \cdots & 0 \\
0 & 0 & \sqrt{2} & 0 & \cdots & 0\\
0 & 0 & 0 & \sqrt{3}& \cdots& 0\\
\vdots & \vdots & \vdots & \vdots & \ddots & \vdots\\
0 & 0 & 0 & 0& \cdots & \sqrt{N-1} \\
0 & 0 & 0 & 0& \cdots & 0
\end{pmatrix}, \\
\hat{b}^\dagger &=
\begin{pmatrix}
0 & 0 & 0 & 0& \cdots & 0 & 0\\
1 & 0 & 0 & 0& \cdots  & 0 & 0\\
0 & \sqrt{2} & 0 & 0& \cdots  & 0 & 0\\
0 & 0 & \sqrt{3} & 0& \cdots  & 0 & 0\\
\vdots & \vdots & \vdots & \vdots& \ddots & \vdots & \vdots \\
0 & 0 & 0 & 0& \cdots  &  \sqrt{N-1}& 0 \\
\end{pmatrix}.
\end{align} 

The operators $\hat b, \hat b^{\dagger}$ are adjoint to each other and satisfy the following algebra:
\begin{equation}
[\hat{b},\hat{b}^{\dagger}] = \begin{pmatrix}
1 & 0 & 0 & 0  & \cdots & 0\\
0 & 1 & 0 & 0& \cdots & 0\\
0 & 0 & 1 & 0 & \cdots & 0\\
\vdots & \vdots & \vdots & \vdots & \ddots & \vdots\\
0 & 0 & 0 & 0 & \cdots & -(N-1) 
\end{pmatrix}
= \hat{\mathbb{I}}-N\ket{N-1}\bra{N-1},
\end{equation}
which differs from the standard infinite-dimensional CCR algebra \eqref{CCRcomms}.
The resulting occupation number operator is:
\begin{equation}
\hat{N}_b:= \hat{b}^{\dagger}\hat{b} =\begin{pmatrix}
0 & 0 & 0 & 0 & \cdots & 0\\
0 & 1 & 0 & 0 & \cdots & 0\\
0 & 0 & 2 & 0 & \cdots & 0\\
0 & 0 & 0 & 3 & \cdots & 0 \\
\vdots & \vdots & \vdots & \vdots & \ddots & \vdots\\
0 & 0 & 0 & 0 & \cdots & N-1 
\end{pmatrix},
\end{equation}
which is symmetric, and the standard eigenvalue equation holds:
\begin{equation}
\hat{N}_b\ket{n} = n \ket{n}.
\end{equation}

Interestingly, the cut-off bosonic operators $\hat b, \hat b^{\dagger}$ follow the same commutation relations with the number operator $\hat N_b$ as in the standard bosonic algebra:
\begin{align}
    \label{comRelbbdN}
    [\hat b,\hat N_b]&=\hat b, \nonumber\\
    [\hat b^{\dagger},\hat N_b]&=-\hat b^{\dagger}.
\end{align}

One can verify that the action of the annihilation operator 
on the eigenstates $\ket{n}$ has the standard form:
\begin{equation}
\hat{b}\ket{n} = \sqrt{n}\ket{n-1},
\end{equation}
for $n\in \{ 0, ..., N-1\}$. On the other hand, the action of the
creation operator is slightly modified. Namely, for 
$n\in \{ 0, ..., N-2\}$ we have:
\begin{equation}
\hat{b}^{\dagger}\ket{n} = \sqrt{n+1}\ket{n+1},
\end{equation}
while for $n=N-1$, the action of $\hat{b}^{\dagger}$ is terminated:
\begin{equation}
\hat{b}^{\dagger}\ket{N-1} = 0.
\end{equation}

\section{Relation to $q$-deformed harmonic oscillator}

The concept of $q$-deformation provides a systematic generalization of canonical
algebraic structures by introducing a deformation parameter $q$ that modifies the
underlying commutation relations. In the context of quantum harmonic oscillators,
$q$-deformation leads to a generalized bosonic algebra that interpolates between the
standard infinite-dimensional case and finite-dimensional representations.

When the deformation parameter \(q\) is chosen to be a root of unity, the corresponding \(q\)-deformed oscillator admits finite-dimensional representations. This provides an algebraic counterpart of the truncated Fock-space construction introduced in the previous section, with the standard bosonic algebra recovered in the limit \(q\to1\).

Here, for the purpose of truncating the resulting Fock space, 
we employ the $q$-deformation of the algebra of creation and 
annihilation operators:
\begin{align}
[\hat{B},\hat{B}^\dagger]_q\equiv\hat{B}\hat{B}^\dagger - q\, \hat{B}^\dagger \hat{B} = \hat{\mathbb{I}},
\end{align}
where $q \in \mathbb{C}$ is the deformation parameter. In 
the $q$-deformed setting, the symbol $\hat B^{\dagger}$ denotes 
a separate operator and should not be treated as a Hermitian
conjugate of $\hat B$.

As explicitly shown in Appendix B, the actions of the $q$-deformed 
annihilation and creation operators are given by:
\begin{align}
\hat{B}\,\ket{n} &= \sqrt{[n]_q}\,\ket{n-1}, \\
\hat{B}^\dagger\,\ket{n} &= \sqrt{[n+1]_q}\,\ket{n+1},
\end{align}
where the $q$-number $[n]_q$ is defined as
\begin{equation}
[n]_q = \frac{1 - q^n}{1 - q}.
\end{equation}
From here one can find, that the $\ket{n}$ state can be represented as 
follows:
\begin{equation}
\ket{n} = \frac{(\hat{B}^{\dagger})^n}{\sqrt{[n]_q!}}\ket{0},
\end{equation}
where $[n]_q! := [1]_q[2]_q...[n]_q$, and we define $[0]_q!=1$.

Furthemore, action of the occupation number operator $\hat{N}_B:= \hat{B}^\dagger\hat{B}$
is:
\begin{align}
\hat{N}_B \ket{n}  = [n]_q \, \ket{n}. 
\end{align}

Here, we considered  the case where $q \in \mathbb{C}$ is the $N$-th root 
of the unity:
\begin{equation}
q^N=1,
\end{equation}
so that:
\begin{equation}
q=e^{2\pi i/N}, 
\end{equation}
and action of the creation operator on the $\ket{N-1}$ state gives zero:
\begin{equation}
\hat{B}^\dagger \ket{N-1}=0. 
\end{equation}

In consequence, the $q$-deformed Fock space is spanned by the 
following states:
\begin{equation}
\mathcal{H}= \text{span} \{ \ket{0},\ket{1},\dots, \ket{N-1} \},
\end{equation}
so that $\dim \mathcal{H} = N$. In what follows, we will focus on 
the special case with $N=2^m$, where $m$ corresponds to the 
number of qubits in the quantum register. 

Introducing the following diagonal operator:
\begin{equation}
\hat{Q}:=\begin{pmatrix}
1 & 0 & 0 & 0 & \cdots & 0\\
0 & q & 0 & 0 & \cdots  & 0\\
0 & 0 & q^2 & 0 & \cdots  & 0\\
0 & 0 & 0 & q^3 & \cdots  & 0 \\
\vdots  & \vdots & \vdots & \vdots & \ddots & \vdots\\
0 & 0 & 0 & 0 & \cdots  & q^{N-1}
\end{pmatrix},
\end{equation}
we can explicitly express the commutation relations for the $\{\hat B,\hat B^{\dagger}\}$ algebra using standard matrix commutators: 
\begin{align}
    \label{comRelBBDN}
    [\hat B,\hat B^{\dagger}]&=\hat Q,\nonumber\\
    [\hat B,\hat N_B]&=\hat Q\hat B,\nonumber\\
    [\hat B^{\dagger},\hat N_B]&=-\hat B^{\dagger}\hat Q.
\end{align}

Please keep in mind that the occupation number operator $\hat{N}_B$ is non-Hermitian in the 
considered finite-dimensional case, with the eigenvalues:
\begin{equation}
[n]_q = e^{i\pi (n-1)/N} \frac{\sin (\pi n/N)}{\sin (\pi/N)}.
\label{qNumberN}
\end{equation}
Now the non-deformed limit ($q\rightarrow 1$) corresponds to 
$N\rightarrow \infty$, so that:
\begin{align}
[n]_q &\rightarrow n, \\
\sqrt{[n]_q} &\rightarrow \sqrt{n}, \\
\sqrt{[n+1]_q} &\rightarrow \sqrt{n+1},
\end{align}
leading to the standard expressions. 

\subsection{$N=2^2=4$}

Let us here discuss the $N=4$ case, in which the Fock space 
is approximated by the four states: $\ket{0}, \ket{1}, \ket{2},$ 
and $\ket{3}$. So, up to three-photon states are included. The 
four states can be represented by $2$-qubit quantum 
register so that $N=2^m=2^2=4$. Furthermore, here $q=i$.   

Here, the occupation number (\ref{qNumberN}) takes the form: 
\begin{equation}
[n]_q = e^{i\pi (n-1)/4} \frac{\sin (\pi n/4)}{\sin (\pi/4)},
\end{equation}
from where 
\begin{align}
[0]_q = 0, \ \ 
[1]_q = 1, \ \
[2]_q = \sqrt{2} e^{i\pi/4} = 1+i, \ \
[3]_q &= i
\end{align}
and 
\begin{align}
[0]_q! &=1, \\ 
[1]_q! &= 1,\\
[2]_q! &= \sqrt{2} e^{i\pi/4} = 1+i,\\
[3]_q! &= \sqrt{2} ie^{i\pi/4} =i-1.
\end{align}

The matrix representations of the annihilation, creation, and 
occupation number operators are as follows:
\begin{align}
\hat{B} &= \begin{pmatrix}
0 & 1 & 0 & 0 \\
0 & 0 & \sqrt[4]{2}e^{i\pi/8} & 0 \\
0 & 0 & 0 & e^{i\pi/4} \\
0 & 0 & 0 & 0 \\
\end{pmatrix},\\
\hat{B}^{\dagger} &=  \begin{pmatrix}
0 & 0 & 0 & 0 \\
1 & 0 & 0 & 0 \\
0 & \sqrt[4]{2}e^{i\pi/8} & 0 & 0 \\
0 & 0 &  e^{i\pi/4} & 0 \\
\end{pmatrix},   \\
\hat{N}_B &=  \begin{pmatrix}
0 & 0 & 0 & 0\\
0 & 1 & 0 & 0\\
0 & 0 & 1+i & 0 \\
0 & 0 & 0 & i \\
\end{pmatrix}. \\
\end{align}

It is clear now that the operators $\hat B$ and $\hat B^{\dagger}$ are \textit{not} adjoint to each other since we have:
\begin{equation}
    \hat B^{\dagger}=(\hat B)^T\neq (\hat B)^{\dagger}, 
\end{equation}
where the last expression means just the ordinary Hermitian conjugate of the matrix $\hat B$. In consequence, the number operator is not Hermitian and has complex eigenvalues. Therefore, the \(q\)-deformed representation considered here should be understood as an algebraic construction rather than as a unitary realization of the truncated optical dynamics. The latter is defined by the mutually adjoint truncated operators \(\hat b\) and \(\hat b^\dagger\) introduced in the previous section.

\subsection{Relation to cut-off Fock space}

The truncated oscillator introduced in the previous section and the present \(q\)-deformed representation can be related by a generally non-unitary diagonal rescaling. This can be done by virtue of the 
rescaling of the creation and annihilation operators
or equivalently, the basis states. The rescaling 
can be made by the following operation:
\begin{equation}
\hat{F}:=\begin{pmatrix}
1 & 0 & 0 & 0 & \cdots & 0\\
0 & \sqrt{\frac{[1]_q}{1}} & 0 & 0 & \cdots  & 0\\
0 & 0 & \sqrt{\frac{[2]_q}{2}} & 0 & \cdots  & 0\\
0 & 0 & 0 & \sqrt{\frac{[3]_q}{3}} & \cdots  & 0 \\
\vdots  & \vdots & \vdots & \vdots & \ddots & \vdots\\
0 & 0 & 0 & 0 & \cdots  & \sqrt{\frac{[N-1]_q}{N-1}}
\end{pmatrix},
\end{equation}
where 
\begin{equation}
q=e^{2\pi i/N}. 
\end{equation}
One can find that the following relations hold:
\begin{align}
\label{Bfromb}
\hat{B}&=\hat{b}\hat{F}, \nonumber\\
\hat{B}^{\dagger}&=\hat{F}\hat{b}^{\dagger}, \nonumber\\
\hat{N}_B &= \hat{F} \hat{N}_b \hat{F}. 
\end{align}
Using the inverse matrix $\hat F^{-1}$ we have:
\begin{align}
\label{bfromB}
\hat{b}&=\hat{B}\hat{F}^{-1}, \nonumber\\
\hat{b}^{\dagger}&=\hat{F}^{-1}\hat{B}^{\dagger}, \nonumber\\
\hat{N}_b &= \hat{F}^{-1} \hat{N}_B \hat{F}^{-1}. 
\end{align}


\section{Mapping convention}
To represent photonic Fock states within the framework of qubit-based quantum circuits, we adopt a binary encoding convention that assigns each photon-number degree of freedom to a register of qubits. This allows for a direct and unambiguous translation between abstract bosonic modes and the computational basis of a standard gate-based quantum computer.

\subsection*{Single-mode encoding}
For a single optical mode with occupation number $n$, we encode the corresponding Fock state $\ket n$ into a register of $m+1$ qubits. The integer $n$ is written in binary form, and each binary digit is stored in one qubit of the register. Throughout this work, we adopt a most-significant-bit (MSB)-first ordering, such that the qubit $q_m$ carries the MSB and $q_0$ the least-significant bit (LSB). The mapping is therefore
\begin{equation}
    \ket{n} = \ket{q_m,\dots q_1q_0},
\end{equation}
where the tuple $(q_m,\dots q_1q_0)$ denotes the binary expansion of $n$.

\begin{figure}[h!]
    \centering
    \begin{quantikz}[baseline=(current bounding box.center)]
        \lstick{$q_m$} & \qw\\
        \lstick{$\vdots$} \\
        \lstick{$q_1$} & \qw\\
        \lstick{$q_0$} & \qw\\
    \end{quantikz}
    \caption{Notation for single-mode quantum circuit.}
    \label{fig:singlemodenotation}
\end{figure}
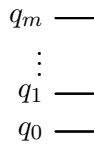

The corresponding quantum circuit representation of the register is Fig. \ref{fig:singlemodenotation}.

This binary encoding is compact and enables efficient manipulation of Fock levels using standard quantum-logic operations.

\subsection*{Multi-mode encoding}
For a multimode photonic system with mode occupations $(n_0,n_1,\dots,n_k)$, we assign an independent qubit register to each mode. Each register is encoded exactly as in the single-mode case, using an identical number of qubits and the same MSB-first ordering. The global Hilbert-space basis state is then obtained by concatenating all mode-specific registers:
\begin{equation}
    \ket{n_0,n_1,\dots,n_k} = \ket{q_m^{(0)},\dots q_1^{(0)}q_0^{(0)}, q_m^{(1)}\dots q_1^{(1)}q_0^{(1)},\dots,q_m^{(k)}\dots q_1^{(k)}q_0^{(k)}}.
\end{equation}

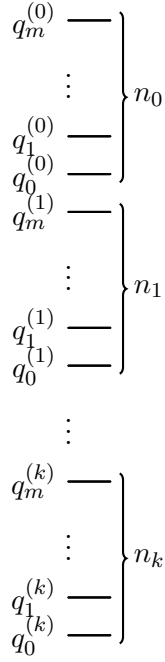
\begin{figure}[h!]
    \centering
    \begin{quantikz}[baseline=(current bounding box.center)]
        \lstick{$q_m^{(0)}$} & \qw\rstick[4]{$n_0$}\\
        \vdots\\
        \lstick{$q_1^{(0)}$} & \qw\\
        \lstick{$q_0^{(0)}$} & \qw\\
        \lstick{$q_m^{(1)}$} & \qw\rstick[4]{$n_1$}\\
        \vdots\\
        \lstick{$q_1^{(1)}$} & \qw\\
        \lstick{$q_0^{(1)}$} & \qw\\
        \vdots\\
        \lstick{$q_m^{(k)}$} & \qw\rstick[4]{$n_k$}\\
        \vdots\\
        \lstick{$q_1^{(k)}$} & \qw\\
        \lstick{$q_0^{(k)}$} & \qw\\
    \end{quantikz}
    \caption{Notation for multi-mode quantum circuit.}
    \label{fig:multimodenotation}
\end{figure}

A schematic circuit depiction of this convention is given below, where groups of $m+1$ qubits correspond to the registers $\{n_j\}$ of individual modes, Fig. \ref{fig:multimodenotation}.

This convention ensures a consistent and scalable embedding of photonic states into qubit-based architectures, enabling the simulation of bosonic systems using standard circuit-model quantum computing tools.


\section{Truncated quantum optics for $N=2$}

In the mode representation, each of the contributing Hilbert spaces, for the $N=2$ case, can be spanned by 
the following basis vectors:
\begin{align}
\ket{0} = \begin{pmatrix} 1 \\ 0 \end{pmatrix},  \ \
\ket{1} = \begin{pmatrix} 0 \\ 1 \end{pmatrix}.
\end{align}
Let us consider the canonised variables for $N=2$, 
so that:
\begin{align}
\hat{b} &=
\begin{pmatrix}
0 & 1 \\
0 & 0 
\end{pmatrix}, \\
\hat{b}^\dagger &=
\begin{pmatrix}
0 & 0 \\
1 & 0 
\end{pmatrix}.
\end{align} 
The operators are hermitian and satisfy the following algebra:
\begin{equation}
[\hat{b},\hat{b}^{\dagger}] = \sigma_z 
= \begin{pmatrix}
1 & 0 \\
0 & -1
\end{pmatrix},
\end{equation}
or the anticommutation relation: 
\begin{equation}
\{\hat{b},\hat{b}^{\dagger}\} = \hat{\mathbb{I}}.
\end{equation}
Therefore, for $N=2$, the truncation corresponds to fermionic statistics. 

\noindent The occupation number operator is:
\begin{equation}
\hat{N}:= \hat{b}^{\dagger}\hat{b} =\begin{pmatrix}
0 & 0 \\
0 & 1 
\end{pmatrix}.
\end{equation}
Since $\hat b,\hat b^\dagger$ are mutually adjoint, all relevant 
single-mode operators follow standard definitions presented below.

\noindent The phase shift operator is:
\begin{equation}
\hat{U}(\theta) := \exp \left(-i\theta \hat{N} \right) = \begin{pmatrix}
1& 0 \\
0 & e^{-i\theta} 
\end{pmatrix}.
\end{equation}

The displacement operator:
\begin{equation}
\hat{D}(\alpha) := \exp \left(\alpha \hat{b}^{\dagger}-\alpha^* \hat{b}\right)
 = \begin{pmatrix}
\cos r & -\,e^{-i\varphi}\,\sin r\\[6pt]
e^{i\varphi}\,\sin r & \cos r
\end{pmatrix},
\end{equation}
where $\alpha=r^{i\varphi}$.

\noindent The squeezing operator reads:
\begin{equation}
\hat{S}(z) := \exp \left[ (z^*\hat{b}^2-z(\hat{b}^{\dagger})^2)/2\right]
= \begin{pmatrix}
1 & 0 \\
0 & 1 
\end{pmatrix}.
\end{equation}
Since \(\hat b^2=(\hat b^\dagger)^2=0\) for \(N=2\), the squeezing operator reduces identically to the identity. Consequently, this lowest-dimensional truncation does not reproduce any nontrivial squeezing dynamics and therefore cannot represent the nonlinear optical processes associated with squeezed-state generation. The \(N=2\) case should thus be regarded primarily as an illustrative example of the encoding and circuit construction rather than as a faithful model of squeezed-light dynamics.

Furthermore, the transformation operator for the general beam splitter can now be expressed as:
\begin{align}
\hat{U}_{BS}^{N=2}(\theta,\phi) &= \exp \left[ \theta\left(e^{i\phi}\hat{b}^{\dagger}_1\otimes\hat{b}_2-e^{-i\phi}\hat{b}_1 \otimes\hat{b}_2^{\dagger}    \right)   \right], \nonumber \\
 &=(\hat{U}(\phi)^\dagger \otimes \hat{\mathbb{I}})
\hat{U}_{BS}^{N=2}(\theta,0)(\hat{U}(\phi)\otimes\hat{\mathbb{I}})
\end{align}
which is unitary. Extraction of the phase $\phi$ dependence applied above
follows the same rule as for the general case in Eq. \ref{UBSDecomposition}.
Then, once write:
\begin{equation}
\hat{U}_{BS}^{N=2}(\theta,0) = e^{\hat{A}},
\end{equation}
where the matrix representation of the $\hat{A}$ operator takes the 
form:
\begin{equation}
\hat{A} = \theta
\begin{pmatrix}
0 & 0 & 0 & 0\\
0 & 0 & -1 & 0\\
0 & 1 & 0 & 0\\
0 & 0 & 0 & 0
\end{pmatrix},
\end{equation}
which is defined in the following basis:
\begin{align}
\ket{0,0}= \begin{pmatrix} 1 \\ 0 \end{pmatrix} \otimes \begin{pmatrix} 1 \\ 0 \end{pmatrix} = \begin{pmatrix} 1 \\ 0 \\ 0 \\ 0 \end{pmatrix},  \ \ \ \
\ket{0,1}= \begin{pmatrix} 1 \\ 0 \end{pmatrix} \otimes \begin{pmatrix} 0 \\ 1 \end{pmatrix} = \begin{pmatrix} 0 \\ 1 \\ 0 \\ 0 \end{pmatrix}, \\
\ket{1,0}= \begin{pmatrix} 0 \\ 1 \end{pmatrix} \otimes \begin{pmatrix} 1 \\ 0 \end{pmatrix} = \begin{pmatrix} 0 \\ 0 \\ 1 \\ 0 \end{pmatrix}, \ \ \ \
\ket{1,1}= \begin{pmatrix} 0 \\ 1 \end{pmatrix} \otimes \begin{pmatrix} 0 \\ 1 \end{pmatrix} = \begin{pmatrix} 0 \\ 0 \\ 0 \\ 1 \end{pmatrix}.
\end{align}
From here:
\begin{equation}
    \hat{U}_{BS}^{N=2}(\theta,0)=e^{\hat{A}}=\left(\begin{matrix}1 & 0 & 0 & 0\\0 & \cos{\left(\theta \right)} & - \sin{\left(\theta \right)} & 0\\0 & \sin{\left(\theta \right)} & \cos{\left(\theta \right)} & 0\\0 & 0 & 0 & 1\end{matrix}\right).
\end{equation}

The number-preserving \(N=2\) interferometric transformations are closely related to fermionic Gaussian, or matchgate, circuits \cite{TerhalDiVincenzo2002,JozsaMiyake2008}.

The transformation can be decomposed according to the total number of 
photons involved, i.e., $n=n_1+n_2$. Then the transformation of the 
basis states can be written as the following direct sum:
\begin{equation}
U^{N=2}_{BS}(\theta,0) = \bigoplus_{i=0}^2 U_i = U_0 \oplus  U_1 \oplus U_2.
\end{equation}
The contributing blocks are:
\begin{itemize}
\item $n=0$ (no photons) - basis ($\ket{0,0}$)
\[
U_{0} =
\begin{pmatrix}
1
\end{pmatrix}.
\]
\item $n=1$ (1 photon) - basis ($\ket{0,1},\ket{1,0}$)
\[
U_{1} = 
\begin{pmatrix}
\cos (\theta) & - \sin (\theta) \\
 \sin (\theta) & \cos (\theta
\end{pmatrix} = R_y(2\theta).
\]
\item $n=2$ (2 photons) - basis ($\ket{1,1}$)
\[
U_{2} =\begin{pmatrix}
1
\end{pmatrix}.
\]
\end{itemize}

\subsection{Quantum gates}

For the case under consideration, the photonic modes 
map $1-1$ to single qubits.  

Consequently, the $\hat{U}(\theta), \hat{D}(\alpha)$ and $\hat{U}^{N=2}_{BS}(\theta,0)$ operators  can now be represented by the following gates:
\begin{align}
\hat{U}(\theta) &= e^{-i\theta/2} R_z(-\theta) = U_1(-\theta),\\
\hat{D}(\alpha) & = R_z(\varphi)\,R_y(2r)\,R_z(-\varphi),\\
\hat{U}^{N=2}_{BS}(\theta,0) &= (R_z(-\pi/2) \otimes \mathbb{I} )U_{XY}(\theta) (R_z(\pi/2) \otimes  \mathbb{I})=G\left(2\theta \right),
\end{align}
where $G$ is the so-called Givens gate. The quantum gates $U_{XY}(\theta)$ 
and $G\left(2\theta \right)$ read:
\begin{eqnarray}
 U_{XY}(\theta)=\exp[i\tfrac{\theta}{2}(X\otimes X+Y\otimes Y)],\nonumber\\
 G(2\theta)=\left(\begin{matrix}1 & 0 & 0 & 0\\0 & \cos{\left(\theta \right)} & - \sin{\left(\theta \right)} & 0\\0 & \sin{\left(\theta \right)} & \cos{\left(\theta \right)} & 0\\0 & 0 & 0 & 1\end{matrix}\right).
\end{eqnarray}

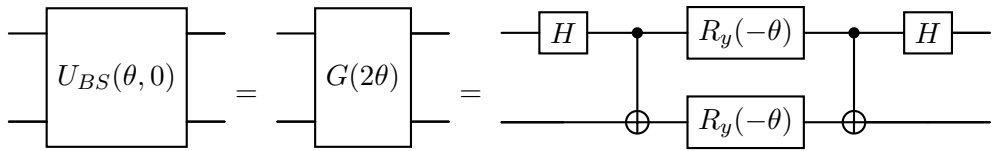
\begin{figure}[h!]
    \centering
    \begin{quantikz}[baseline=(current bounding box.center), transparent]
        \qw & \gate[2]{U_{BS}(\theta,0)} & \qw\\
        \qw & \qw& \qw\\
    \end{quantikz}
    =
    \begin{quantikz}[baseline=(current bounding box.center), transparent]
        \qw & \gate[2]{G(2\theta)} & \qw\\
        \qw & \qw& \qw\\
    \end{quantikz}
    =
    \begin{quantikz}[baseline=(current bounding box.center), transparent]
        \qw & \gate{H} & \ctrl{1} & \gate{R_y(-\theta)} & \ctrl{1} & \gate{H} & \qw\\
        \qw & \qw      & \targ{}  & \gate{R_y(-\theta)} & \targ{}  & \qw      & \qw\\
    \end{quantikz}
    \caption{Circuit for $\hat U_{BS}(\theta,0)$ for $N=2$.}
    \label{fig:givens_circ}
\end{figure}

Fig. \ref{fig:givens_circ} shows the decomposition of the Givens operator involved. 

\subsection{Example 1: BS for two coherent states with $N=2$}

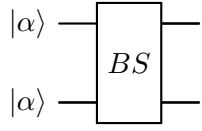
\begin{figure}[ht!]
    \centering
    \begin{quantikz}
        \lstick{\ket{\alpha}} 
            & \gate[2]{BS} 
            & \qw \\
        \lstick{\ket{\alpha}} 
            & \qw 
            & \qw
    \end{quantikz}
    \caption{Exemplary optical circuit with symmetric beam splitter BS ($U_{BS}(\pi/4),0$).}
    \label{fig:2DBS}
\end{figure}

As an example, let us consider the optical circuit shown in Fig.  \ref{fig:2DBS}, where $\alpha=re^{i\varphi}$.

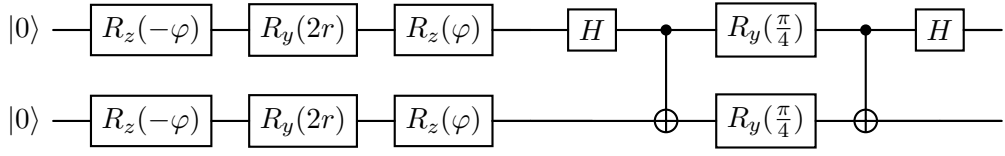
\begin{figure}[ht!]
\centering
    \begin{quantikz}[baseline=(current bounding box.center), transparent]
     \lstick{\ket{0}}  \qw & \gate{R_z(-\varphi)} & \gate{R_y(2r)} & \gate{R_z(\varphi)} & \qw & \gate{H} & \ctrl{1} & \gate{R_y(\frac{\pi}{4})} & \ctrl{1} & \gate{H} & \qw\\
       \lstick{\ket{0}} \qw & \gate{R_z(-\varphi)} & \gate{R_y(2r)} & \gate{R_z(\varphi)} & \qw & \qw      & \targ{}  & \gate{R_y(\frac{\pi}{4})} & \targ{}  & \qw      & \qw\\
    \end{quantikz}
    \caption{Quantum circuit corresponding to the optical circuit (\ref{fig:2DBS}) for $N=2$.}
    \label{fig:2DBSqubit}
\end{figure}

The corresponding circuit in the 2-qubit register for $N=2$ truncation 
in shown in Fig. \ref{fig:2DBSqubit}.

\subsection{Example 2: Four optical modes with $N=2$}

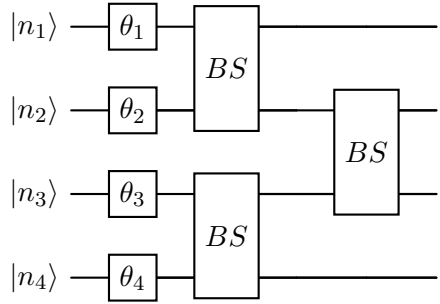
\begin{figure}[ht!]
    \centering
    \begin{quantikz}
        \lstick{\ket{n_1}} 
           & \gate{\theta_1} & \gate[2]{BS} & \qw & \qw & \qw\\
        \lstick{\ket{n_2}} 
           & \gate{\theta_2} & \qw & \qw & \gate[2]{BS} & \qw\\
        \lstick{\ket{n_3}} 
           & \gate{\theta_3} & \gate[2]{BS} & \qw & \qw & \qw\\
        \lstick{\ket{n_4}} 
           & \gate{\theta_4} & \qw & \qw & \qw & \qw
    \end{quantikz}
    \caption{Exemplary optical circuit with four optical modes.}
    \label{fig:4BS}
\end{figure}

Fig. \ref{fig:4BS} shows a simple 4-mode optical circuit which can 
be regarded as a prototype of the boson sampling circuit. Here, the $BS$ 
corresponds to symmetric beam splitters, for which $(\theta,\phi)=(\pi/4,0)$. 
However, please notice that for the truncation under consideration 
($N=2$) the circuit provides a basic example of fermionic sampling. 

\begin{figure}[ht!]
    \centering
    \begin{quantikz}
        \lstick{\ket{0}} 
           & \gate{X^{n_1}} & \gate{U_1(-\theta_1)} & \gate[2]{G(\frac{\pi}{2})} & \qw & \qw & \qw\\
        \lstick{\ket{0}} 
          & \gate{X^{n_2}} & \gate{U_1(-\theta_2)} & \qw & \qw & \gate[2]{G(\frac{\pi}{2})} & \qw\\
        \lstick{\ket{0}} 
           & \gate{X^{n_3}}& \gate{U_1(-\theta_3)} & \gate[2]{G(\frac{\pi}{2})} & \qw & \qw & \qw\\
        \lstick{\ket{0}} 
          & \gate{X^{n_4}} & \gate{U_1(-\theta_4)} & \qw & \qw & \qw & \qw
    \end{quantikz}
    \caption{Quantum circuit corresponding to the optical circuit (\ref{fig:4BS}) for $N=2$.}
    \label{fig:2DBQ}
\end{figure}
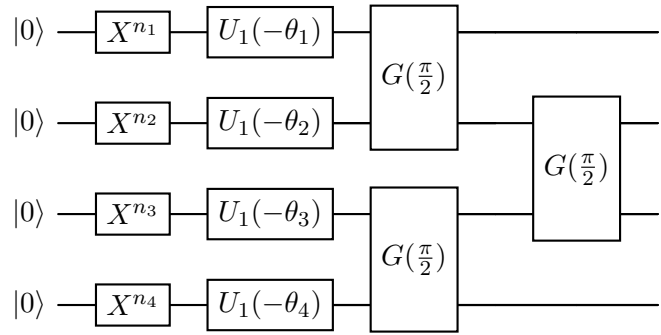

Fig. \ref{fig:2DBQ} shows a qubit-based circuit corresponding to 
the optical circuit shown in Fig. \ref{fig:4BS}.

\section{Truncated quantum optics for $N=4$}

In the mode representation, the Fock space, for the $N=4$ case, 
can be spanned by the following basis vectors:
\begin{align}
\ket{0} = \begin{pmatrix} 1 \\ 0 \\ 0 \\ 0 \end{pmatrix},  \ \
\ket{1} = \begin{pmatrix} 0 \\ 1 \\ 0 \\ 0 \end{pmatrix},  \ \
\ket{2} = \begin{pmatrix} 0 \\ 0 \\ 1 \\ 0 \end{pmatrix},  \ \
\ket{3} = \begin{pmatrix} 0 \\ 0 \\ 0 \\ 1 \end{pmatrix}.  \ \
\end{align}
The truncated creation and annihilation operators for $N=4$ read:
\begin{align}
\label{b4}
\hat{b} &=
\begin{pmatrix}
0 & 1 & 0 & 0\\
0 & 0 & \sqrt{2} & 0\\
0 & 0 & 0 & \sqrt{3}\\
0 & 0 & 0 & 0
\end{pmatrix}, \\
\hat{b}^\dagger &=
\begin{pmatrix}
0 & 0 & 0 & 0\\
1 & 0 & 0 & 0\\
0 & \sqrt{2} & 0 & 0\\
0 & 0 & \sqrt{3} & 0
\end{pmatrix}.
\end{align} 
The operators are adjoint to each other and satisfy the following algebra:
\begin{equation}
[\hat{b},\hat{b}^{\dagger}] = \begin{pmatrix}
1 & 0 & 0 & 0\\
0 & 1 & 0 & 0\\
0 & 0 & 1 & 0\\
0 & 0 & 0 & -3
\end{pmatrix}.
\end{equation}
The occupation number operator is:
\begin{equation}
\hat{N}:= \hat{b}^{\dagger}\hat{b} =\begin{pmatrix}
0 & 0 & 0 & 0\\
0 & 1 & 0 & 0\\
0 & 0 & 2 & 0\\
0 & 0 & 0 & 3
\end{pmatrix}.
\end{equation}
Since the operators $\hat b,\hat b^\dagger$ are mutually adjoint, 
hence $\hat N$ is Hermitian and the exponents are anti-Hermitian, 
all the relevant single-mode operators can be introduced in the 
standard way:
\begin{align}
\hat{D}(\alpha) & := \exp \left(\alpha \hat{b}^{\dagger}-\alpha^* \hat{b}\right), \\
\hat{S}(z) & := \exp \left[ (z^*\hat{b}^2-z(\hat{b}^{\dagger})^2)/2\right], \\
\hat{U}(\theta) & := \exp \left(-i\theta \hat{N} \right).
\end{align}
The operators are unitary, since the exponents 
can be proven anti-Hermitian. 
\\

\subsection{Phase shift operator}

First, observe that for $n$ qubits, the number operator 
$\hat{N}^{(n)}$ can be written as:
\begin{equation}\label{eq:N_general}
    \hat{N}^{(n)} = \sum_{j=0}^{n-1}2^j \hat{N}_j,
\end{equation}
where 
\begin{equation}
    \hat{N}_j = \underbrace{\hat{\mathbb{I}}}_{(n-1)^{th}}\otimes \dots \otimes \underbrace{\frac{\hat{\mathbb{I}}-\hat{Z}}{2}}_{j^{th}}\otimes \dots \otimes \underbrace{\hat{\mathbb{I}}}_{0^{th}}
\end{equation}
 is the local operator that reads the $j^{th}$ qubit value in the binary basis encoding scheme,
 \begin{equation}
     \frac{\hat{\mathbb{I}}-\hat{Z}}{2} |0\rangle = 0 | 0 \rangle, \ \ \ \ \frac{\hat{\mathbb{I}}-\hat{Z}}{2} |1\rangle = 1 | 1 \rangle.
 \end{equation}
 With the $2^j$ factors in the formula~(\ref{eq:N_general}), we decode the binary occupation number, and we obtain a general number operator $\hat{N}^{(n)}$.
 Therefore, the general $n$ qubit phase shift operator can be described with local $\hat{R}_Z$ rotations
 \begin{align}
     \exp\left( -i \theta \hat{N}^{(n)} \right) &= \exp\left(-i\theta \frac{2^n-1}{2}\right) \prod_{j=0}^{n-1}\exp\left( i\frac{\theta}{2}2^j \hat{Z}_j \right) = \bigg| \hat{R}_z(\theta) = \exp\left(-\frac{\theta}{2}\hat{Z}\right)  \bigg|=\\
     &= \exp\left(-i\theta \frac{2^n-1}{2}\right) \prod_{j=0}^{n-1}\hat{R}_{z,j}(-2^j \theta) = \hat{U}(\theta).
 \end{align}

For the case under consideration, we find that the phase shift operator is:
\begin{align}
\hat{U}(\theta)  = \begin{pmatrix}
1 & 0 & 0 & 0\\
0 & e^{-i\theta} & 0 & 0\\
0 & 0 &  e^{-i2\theta} & 0\\
0 & 0 & 0 &  e^{-i3\theta}
\end{pmatrix} = \exp(-i\theta \hat{N}),\\
\end{align}
where $\hat{N} = \text{diag}(0,1,2,3)$. It is a diagonal operator that can be expressed through one-qubit diagonal operations: $\mathbb{I}_2=\text{diag}(1,1)$ and $Z_2=\text{diag}(1,-1)$. Observe that,
\begin{equation}
    \hat{N} = \frac{3}{2} \mathbb{I}_2 \otimes \mathbb{I}_2 - (Z_2 \otimes \mathbb{I}_2) - \frac{1}{2}(\mathbb{I}_2 \otimes Z_2).
\end{equation}
Thus, with all the terms commuting we have,
\begin{align}
    \hat{U}(\theta) &= \exp \left(-\frac{3}{2}i\theta \right) \exp \left(i\theta (Z_2\otimes \mathbb{I}_2)\right) \exp \left(\frac{1}{2}i\theta (\mathbb{I}_2 \otimes Z_2)\right)\\
    &= \bigg| R_z(\alpha) = \exp \left(-i\frac{\alpha}{2} Z\right) \bigg| = \exp \left(-\frac{3}{2}i\theta \right) R_z^{(0)}(-2\theta) R_z^{(1)}(-\theta).
\end{align}

The global phase in the above expression can be discarded, 
leading to the following gate decomposition:
\begin{center}
\begin{quantikz}[baseline=(current bounding box.center)]
\lstick{$q_0$} & \gate[2]{\hat{U}^{4}(\theta)} & \qw\\
\lstick{$q_1$} &                                & \qw
\end{quantikz}
$\equiv$
\begin{quantikz}[baseline=(current bounding box.center)]
 \lstick{$q_0$} & \gate{R_z(-2\theta)} & \qw \\
 \lstick{$q_1$} & \gate{R_z(-\theta)} & \qw       
\end{quantikz}
\end{center}

\subsection{Displacement operator}

The displacement operator is:
\begin{equation}
    \hat{D}(\alpha) = \exp\left(\alpha \hat{b}^{\dagger}- \alpha^* \hat{b}\right) = \exp\left(r[e^{i\phi}\hat{b}^{\dagger} - e^{-i\phi}b]\right),
\end{equation}
where $\alpha=r\,e^{i\phi}$. 
Following an analogous procedure as for the derivation of 
Eq. \ref{DDecomposition}, we find that the displacement 
operator can be expressed in the following form:
\begin{align}
    \hat{D}(\alpha) = \hat{U}^{\dagger}(\phi) \hat{D}(|\alpha|) \hat{U}(\phi).
\end{align}

The $\hat{D}(|\alpha|)$ operator is given by the following matrix:
\begin{equation}
\scalebox{0.6}{$
\hat{D}(|\alpha|)=
\left(
\begin{array}{cccc}
 \frac{1}{6} \left(\left(\sqrt{6}+3\right) c_--\left(\sqrt{6}-3\right) c_+\right) & -\frac{1}{2} \left(s_-+s_+\right) & \frac{\sqrt{3}}{6}(c_--c_+) & \frac{1}{2} \left(s_+-s_-\right) \\
 \frac{1}{2} \left(s_-+s_+\right) & \frac{1}{2} \left(c_-+c_+\right) & \frac{1}{2} \left((\sqrt{3}-\sqrt{2}) s_--\left(\sqrt{2}+\sqrt{3}\right) s_+\right) & \frac{1}{2} \left(c_--c_+\right) \\
 \frac{\sqrt{3}}{6}(c_--c_+) & \frac{1}{2} \left(\left(\sqrt{2}-\sqrt{3}\right) s_-+\left(\sqrt{2}+\sqrt{3}\right) s_+\right) & \frac{1}{6} \left(\left(\sqrt{6}+3\right) c_+-\left(\sqrt{6}-3\right) c_-\right) & \frac{1}{2} \left(\left(\sqrt{2}-\sqrt{3}\right) s_--\left(\sqrt{2}+\sqrt{3}\right) s_+\right) \\
 \frac{1}{2} \left(s_--s_+\right) & \frac{1}{2} \left(c_--c_+\right) & \frac{1}{2} \left((\sqrt{3}-\sqrt{2}) s_-+\left(\sqrt{2}+\sqrt{3}\right) s_+\right) & \frac{1}{2} \left(c_-+c_+\right) \\
\end{array}
\right),
$}
\end{equation}
where 
\[
\omega_\pm=\sqrt{\,3\pm\sqrt6\,},\quad
c_\pm=\cos(\omega_\pm r),\quad
s_\pm=\frac{\sin(\omega_\pm r)}{\omega_\pm }.
\]

Furthermore, by defining $C_\pm=c_-\pm c_+$ and $S_\pm=s_-\pm s_+$, 
the matrix can be further simplified to:
\begin{equation}
\hat{D}(|\alpha|)=\frac{1}{2}
\left(
\begin{array}{cccc}
  \frac{\sqrt{6}}{3}C_-+C_+ & - S_+ & \frac{\sqrt{3}}{3}C_- &  -S_- \\
 S_+ &  C_+ &  \sqrt{3}S_--\sqrt{2}S_+& C_- \\
 \frac{\sqrt{3}}{3}C_- &  \sqrt{2}S_+-\sqrt{3}S_- & -\frac{\sqrt{6}}{3}C_-+C_+ & \sqrt{2}S_--\sqrt{3}S_+ \\
  S_- &  C_- &\sqrt{3}S_+-\sqrt{2}S_- & C_+ \\
\end{array}
\right).
\end{equation}

The magic-basis representation provides the standard connection between local two-qubit operations and \(SO(4)\) transformations \cite{HillWootters1997,KrausCirac2001,Makhlin2002,Vatan:2004nmz}. Namely: 
\begin{equation}
    W = MDM^\dagger=A\otimes B,\quad A,B\in SU(2),
\end{equation}
where
\begin{equation}
    M = \frac{1}{\sqrt2}\begin{pmatrix}
        1 & i & 0 & 0\\
        0 & 0 & i & 1\\
        0 & 0 & i & -1\\
        1 & -i & 0 & 0\\
    \end{pmatrix}
    \label{eq:magic_matrix},
\end{equation}
which can be expressed in terms of gates:
\begin{figure}[ht!]
    \centering
    \begin{quantikz}[baseline=(current bounding box.center)]
        \qw & \gate{S} & \qw & \targ{} & \qw\\
        \qw & \gate{S} & \gate{H} & \ctrl{-1} & \qw\\
    \end{quantikz}
\end{figure}

Therefore, the matrix $D\in SO(4)$ can be expressed as a circuit:
\begin{figure}[ht!]
    \centering
    \begin{quantikz}[baseline=(current bounding box.center)]
        \qw & \gate{S} & \qw & \targ{} & \gate{A} & \targ{} & \qw & \gate{S^\dagger} & \qw\\
        \qw & \gate{S} & \gate{H} & \ctrl{-1} & \gate{B} & \ctrl{-1} & \gate{H} & \gate{S^\dagger} & \qw\\
    \end{quantikz}
\end{figure}

The explicit form of the $W$ matrix is as follows:
\begin{equation}
\scalebox{0.6}{$
    W = 
    \frac{1}{4}
    \begin{pmatrix}2 C_{+} + \frac{\sqrt{6} }{3}C_{-} + 2 i S_{+} & C_{-} \left(\frac{-\sqrt{3}+3}{3}\right)i - \sqrt{2} S_{+} + S_{-} \left(1 + \sqrt{3}\right) & -C_{-} \left(\frac{\sqrt{3}+3}{3}\right)i - \sqrt{2} S_{+} + S_{-} \left(-1 + \sqrt{3}\right) & \frac{\sqrt{6}}{3}C_{-}\\
    C_{-} \left(\frac{\sqrt{3} -3}{3}\right)i + \sqrt{2} S_{+} + S_{-} \left(1 - \sqrt{3}\right) & 2 C_{+} - \frac{\sqrt{6}}{3}C_{-} - 2 \sqrt{3} i S_{+} + 2 \sqrt{2} i S_{-} & - \frac{\sqrt{6}}{3}C_{-} & C_{-} \left(\frac{\sqrt{3}+3}{3}\right) - \sqrt{2} S_{+} + S_{-} \left(1 + \sqrt{3}\right)\\
    C_{-} \left(\frac{\sqrt{3} i}{3} + i\right) + \sqrt{2} S_{+} + S_{-} \left(- \sqrt{3} - 1\right) & - \frac{\sqrt{6}}{3}C_{-} & 2 C_{+} - \frac{\sqrt{6} }{3}C_{-} + 2 \sqrt{3} i S_{+} - 2 \sqrt{2} i S_{-} & C_{-} \left(\frac{\sqrt{3} -3}{3}\right) - \sqrt{2} S_{+} + S_{-} \left(-1 + \sqrt{3}\right)\\
    \frac{\sqrt{6}}{3}C_{-} &- C_{-} \left(\frac{\sqrt{3}+3}{3}\right) + \sqrt{2} S_{+} + S_{-} \left(1 - \sqrt{3}\right) & C_{-} \left(\frac{-\sqrt{3} +3}{3}\right) + \sqrt{2} S_{+} + S_{-} \left(- \sqrt{3} - 1\right) & 2 C_{+} + \frac{\sqrt{6} }{3}C_{-} - 2 i S_{+}\end{pmatrix}$}.
\end{equation}

Consider a general decomposition of some tensor product matrix $W$. We can decompose it into $W=\tilde{A}\otimes\tilde{B}$ where we define:
\begin{equation}
    \tilde A \equiv \begin{pmatrix}
        W_{00} & W_{02}\\
        W_{20} & W_{22}\\
    \end{pmatrix},
\end{equation}
and 
\begin{equation}
    \tilde B \equiv \frac{1}{W_{00}}\begin{pmatrix}
        W_{00} & W_{01}\\
        W_{10} & W_{11}\\
    \end{pmatrix}.
\end{equation}
It is easy to check that:
\begin{equation}
    \tilde A = \begin{pmatrix}
        A_{00}B_{00} & A_{01}B_{00}\\
        A_{10}B_{00} & A_{11}B_{00}\\
    \end{pmatrix}
    = B_{00}A,
\end{equation}
and
\begin{equation}
    \tilde B = \frac{1}{A_{00}B_{00}}\begin{pmatrix}
        A_{00}B_{00} & A_{00}B_{01}\\
        A_{00}B_{10} & A_{00}B_{11}\\
    \end{pmatrix}
    = \frac{1}{B_{00}}B,
\end{equation}
where $A_{ij}$ and $B_{ij}$ are matrix elements of $A$ and $B$. So:
\begin{equation}
    \tilde A \otimes \tilde B = A\otimes B = W.
\end{equation}
The matrices $\tilde A$ and $\tilde B$ give a proper tensor product but are not $SU(2)$ matrices, we need to normalize them by dividing by $\sqrt{\det}$, then we obtain $A,B\in SU(2)$ which we are looking for.
\begin{align}
    A &= \frac{\tilde A}{\sqrt{\det \tilde A}}, \\
    B &= \frac{\tilde B}{\sqrt{\det \tilde B}}.
\end{align}

It is easy to check the above equalities, noticing that: $\det\tilde A=B_{00}^2\det A, \;\det\tilde B=\frac{1}{B_{00}^2}\det B$ and $\det A=\det B=1$.

Applying above discussion to our case, we have:
\begin{equation}
    \tilde A = \frac{1}{4} \begin{pmatrix}
        2 C_{+} + \frac{\sqrt{6} }{3}C_{-} + 2 i S_{+} &  -C_{-} \left(\frac{\sqrt{3}+3}{3}\right)i - \sqrt{2} S_{+} + S_{-} \left(-1 + \sqrt{3}\right) \\
        C_{-} \left(\frac{\sqrt{3} i}{3} + i\right) + \sqrt{2} S_{+} + S_{-} \left(- \sqrt{3} - 1\right) & 2 C_{+} - \frac{\sqrt{6} }{3}C_{-} + 2 \sqrt{3} i S_{+} - 2 \sqrt{2} i S_{-} \\
    \end{pmatrix},
\end{equation}
and
\begin{equation}
    \tilde B = \frac{1}{2 C_{+} + \frac{\sqrt{6} }{3}C_{-} + 2 i S_{+}}\begin{pmatrix}
        2 C_{+} + \frac{\sqrt{6} }{3}C_{-} + 2 i S_{+} &  C_{-} \left(\frac{-\sqrt{3}+3}{3}\right)i - \sqrt{2} S_{+} + S_{-} \left(1 + \sqrt{3}\right)\\
        C_{-} \left(\frac{\sqrt{3} -3}{3}\right)i + \sqrt{2} S_{+} + S_{-} \left(1 - \sqrt{3}\right) & 2 C_{+} - \frac{\sqrt{6}}{3}C_{-} - 2 \sqrt{3} i S_{+} + 2 \sqrt{2} i S_{-}
    \end{pmatrix},
\end{equation}
then:
\begin{align}
    \det\tilde A &=\frac{C_{+}^{2}}{4} + C_{+} S_{+} \left(\frac{i}{4} + \frac{\sqrt{3} i}{4}\right) - \frac{\sqrt{2} i C_{+} S_{-}}{4} + C_{-}^{2} \left(- \frac{1}{8} - \frac{\sqrt{3}}{24}\right) + \frac{\sqrt{2} i C_{-} S_{+}}{4} \nonumber \\
    &+ C_{-} S_{-} \left(- \frac{5 \sqrt{3} i}{24} - \frac{i}{8}\right) 
    + S_{+}^{2} \left(\frac{1}{8} - \frac{\sqrt{3}}{4}\right) + S_{+} S_{-} \left(- \frac{\sqrt{6}}{8} + \frac{\sqrt{2}}{4}\right) + \frac{S_{-}^{2}}{8},
\end{align}
and 
\begin{align}
\nonumber
\det\tilde B 
&=
\frac{3}{%
\begin{aligned}
&36 C_{+}^{3}
+ 18 \sqrt{6} C_{+}^{2} C_{-}
+ 108 i C_{+}^{2} S_{+}
+ 18 C_{+} C_{-}^{2} \\
&+ 36 \sqrt{6} i C_{+} C_{-} S_{+}
- 108 C_{+} S_{+}^{2}
+ \sqrt{6} C_{-}^{3}
+ 18 i C_{-}^{2} S_{+} \\
&- 18 \sqrt{6} C_{-} S_{+}^{2}
- 36 i S_{+}^{3}
\end{aligned}
}
\\[1ex]
&\quad \times
\begin{aligned}
\Bigl(&
12 C_{+}^{3}
+ 2 \sqrt{6} C_{+}^{2} C_{-}
- 12 \sqrt{3} i C_{+}^{2} S_{+}
+ 24 i C_{+}^{2} S_{+} \\
&+ 12 \sqrt{2} i C_{+}^{2} S_{-}
- 6 C_{+} C_{-}^{2}
+ 2 \sqrt{3} C_{+} C_{-}^{2}
- 18 \sqrt{2} i C_{+} C_{-} S_{+} \\
&+ 2 \sqrt{6} i C_{+} C_{-} S_{+}
- 6 i C_{+} C_{-} S_{-}
+ 14 \sqrt{3} i C_{+} C_{-} S_{-} \\
&- 6 C_{+} S_{+}^{2}
+ 24 \sqrt{3} C_{+} S_{+}^{2}
- 24 \sqrt{2} C_{+} S_{+} S_{-}
- 6 \sqrt{6} C_{+} S_{+} S_{-} \\
&+ 6 C_{+} S_{-}^{2}
- \sqrt{6} C_{-}^{3}
+ \sqrt{2} C_{-}^{3}
- 6 i C_{-}^{2} S_{+}
- 2 \sqrt{3} i C_{-}^{2} S_{+} \\
&- \sqrt{6} i C_{-}^{2} S_{-}
+ 5 \sqrt{2} i C_{-}^{2} S_{-}
+ \sqrt{6} C_{-} S_{+}^{2}
+ 18 \sqrt{2} C_{-} S_{+}^{2} \\
&- 14 \sqrt{3} C_{-} S_{+} S_{-}
+ \sqrt{6} C_{-} S_{-}^{2}
+ 6 i S_{+}^{3}
+ 12 \sqrt{3} i S_{+}^{3} \\
&- 12 \sqrt{2} i S_{+}^{2} S_{-}
- 6 \sqrt{6} i S_{+}^{2} S_{-}
+ 6 i S_{+} S_{-}^{2}
\Bigr).
\end{aligned}
\end{align}

\subsection{Squeezing operator}

The single-mode squeezing operator is defined as:
\begin{equation}
\hat{S}(z)= \exp \left[ (z^*\hat{b}^2-z (\hat{b}^{\dagger})^2)/2\right],
\end{equation}
where the complex squeezing parameter is typically represented as $z=re^{i\phi}$.

The truncated squeezing operator can be expressed as follows:
\begin{align}
\label{sq4}
\hat S(z)= 
\left(
\begin{array}{cccc}
 \cos \left(\frac{r}{\sqrt{2}}\right) & 0 & e^{-i \phi } \sin \left(\frac{r}{\sqrt{2}}\right) & 0 \\
 0 & \cos \left(\sqrt{\frac{3}{2}} r\right) & 0 & e^{-i \phi } \sin \left(\sqrt{\frac{3}{2}} r\right) \\
 -e^{i \phi } \sin \left(\frac{r}{\sqrt{2}}\right) & 0 & \cos \left(\frac{r}{\sqrt{2}}\right) & 0 \\
 0 & -e^{i \phi } \sin \left(\sqrt{\frac{3}{2}} r\right) & 0 & \cos \left(\sqrt{\frac{3}{2}} r\right) \\
\end{array}
\right).
\end{align}

This operator has specific structure, changing base using $SWAP$ gate we obtain diagonal-block structure:
\begin{align}
    S' &=SWAP\cdot S\cdot SWAP^\dagger \nonumber \\
       &=\begin{pmatrix}
        \cos \left(\frac{r}{\sqrt{2}}\right) & e^{-i \phi } \sin \left(\frac{r}{\sqrt{2}}\right) & 0& 0\\
        -e^{i \phi } \sin \left(\frac{r}{\sqrt{2}}\right) & \cos \left(\frac{r}{\sqrt{2}}\right)& 0& 0\\
        0& 0&\cos \left(\sqrt{\frac{3}{2}} r\right)&e^{-i \phi } \sin \left(\sqrt{\frac{3}{2}} r\right) \\
        0& 0&-e^{i \phi } \sin \left(\sqrt{\frac{3}{2}} r\right)&\cos \left(\sqrt{\frac{3}{2}} r\right)
    \end{pmatrix}.
\end{align}
Each block is a $U_3$ gate and can be further decomposed, where $\theta=\sqrt2r$ and $\theta'=\sqrt6r$:
\begin{align}
    S'&=\left(\begin{array}{c|c}
        U_3(\theta,\pi+\phi,\pi-\phi)&0  \\
        \hline
        0&U_3(\theta',\pi+\phi,\pi-\phi)\\
    \end{array}\right) \nonumber \\ &=
    \left(\begin{array}{c|c}
        R_z(\phi-\pi)R_y(\theta)R_z(\pi-\phi)&0\\
        \hline
        0&R_z(\phi-\pi)R_y(\theta')R_z(\pi-\phi)\\
    \end{array}\right) \nonumber \\ \nonumber&=
    \left(\begin{array}{c|c}
        R_z(\phi-\pi)&0\\
        \hline
        0&R_z(\phi-\pi)\\
    \end{array}\right)
    \left(\begin{array}{c|c}
        R_y(\theta)&0\\
        \hline
        0&R_y(\theta')\\
    \end{array}\right)
    \left(\begin{array}{c|c}
        R_z(\pi-\phi)&0\\
        \hline
        0&R_z(\pi-\phi)\\
    \end{array}\right)\\ \nonumber&=
    (\mathbb{I}\otimes R_z(\phi-\pi))
    \left(\begin{array}{c|c}
        R_y(\theta)&0\\
        \hline
        0&R_y(\theta)\\
    \end{array}\right)
    \left(\begin{array}{c|c}
        \mathbb{I}&0\\
        \hline
        0&R_y(\theta'-\theta)\\
    \end{array}\right)
    (\mathbb{I}\otimes R_z(\pi-\phi))\\ &=
    (\mathbb{I}\otimes R_z(\phi-\pi))
    (\mathbb{I}\otimes R_y(\theta))
    CR_y(\theta'-\theta)
    (\mathbb{I}\otimes R_z(\pi-\phi)).
\end{align}

Gathering all together we have circuit:
\begin{center}
\begin{quantikz}[baseline=(current bounding box.center)]
& \gate[2]{\hat{S}(z)} & \qw\\
&                      & \qw
\end{quantikz}
$\equiv$
\begin{quantikz}[baseline=(current bounding box.center)]
& \gate{R_z(\pi-\phi)} & \gate{R_y(\theta'-\theta)} & \gate{R_y(\theta)} &\gate{R_z(\phi-\pi)} & \qw\\
& \qw & \ctrl{-1} & \qw & \qw& \qw
\end{quantikz}
\end{center}

Decomposing $CR_y$ gate:
\begin{center}
\begin{quantikz}[baseline=(current bounding box.center)]
& \ctrl{1} & \qw\\
&  \gate{R_y(\theta)}     & \qw
\end{quantikz}
$\equiv$
\begin{quantikz}[baseline=(current bounding box.center)]
&\ctrl{1}&\qw&\ctrl{1}&\qw&\qw\\
&\targ{}&\gate{R_y(-\frac{\theta}{2})}&\targ{}&\gate{R_y(\frac{\theta}{2})}&\qw
\end{quantikz}
\end{center}
we obtain the final circuit for the squeezing operator:
\begin{center}
\begin{quantikz}[baseline=(current bounding box.center)]
& \gate[2]{\hat{S}(z)} & \qw\\
&                      & \qw
\end{quantikz}
$\equiv$
\begin{quantikz}[baseline=(current bounding box.center)]
& \gate{R_z(\pi-\phi)} & \targ{} & \gate{R_y(\frac{\theta-\theta'}{2})}&\targ{}& \gate{R_y(\frac{\theta'+\theta}{2})} &\gate{R_z(\phi-\pi)}& \qw \\
& \qw & \ctrl{-1} & \qw &\ctrl{-1}& \qw& \qw& \qw
\end{quantikz}
\end{center}

\subsection{Beam Splitter}

Finally, the transformation operator for the symmetric beam splitter 
can now be expressed as:
\begin{equation}
\hat{U}_{BS}= \exp \left[ \theta\left(e^{i\phi}\hat{b}^{\dagger}_1\otimes\hat{b}_2-e^{-i\phi}\hat{b}_1\otimes \hat{b}_2^{\dagger}    \right)   \right],
\end{equation}
where the phase $\phi$ dependence can be extracted as follows:
\begin{equation}
\hat{U}_{BS}(\theta,\phi)=(\hat{U}(\phi)^\dagger \otimes\hat{\mathbb{I}})
\hat{U}_{BS}(\theta,0)(\hat{U}(\phi)\otimes\hat{\mathbb{I}}),
\end{equation}
so that 
\begin{equation}
\hat{U}_{BS}(\theta,0) = e^{\hat{A}}.
\end{equation}

The action of this operator is on the tensor product of two truncated
Hilbert spaces, so that photonic basis states $\ket{n_1,n_2}$ in the
product state can be written as:
\begin{align}
\ket{n_1,n_2} = \begin{pmatrix} \delta_{n_1,0} \\ \delta_{n_1,1} \\ \delta_{n_1,2} \\ \delta_{n_1,3} \end{pmatrix} 
\otimes
\begin{pmatrix} \delta_{n_2,0} \\ \delta_{n_2,1} \\ \delta_{n_2,2} \\ \delta_{n_2,3} \end{pmatrix} 
= \begin{pmatrix} 
\delta_{n_1,0}\delta_{n_2,0}  \\ \delta_{n_1,0}\delta_{n_2,1}  \\ \delta_{n_1,0}\delta_{n_2,2}  \\ \delta_{n_1,0}\delta_{n_2,3}    \\ 
\delta_{n_1,1}\delta_{n_2,0}  \\ \delta_{n_1,1}\delta_{n_2,1} \\ \delta_{n_1,1}\delta_{n_2,2}   \\ \delta_{n_1,1}\delta_{n_2,3}    \\ 
\delta_{n_1,2}\delta_{n_2,0}  \\ \delta_{n_1,2}\delta_{n_2,1}  \\ \delta_{n_1,2}\delta_{n_2,2}   \\ \delta_{n_1,2}\delta_{n_2,3}   \\ 
\delta_{n_1,3}\delta_{n_2,0}  \\ \delta_{n_1,3}\delta_{n_2,1}  \\ \delta_{n_1,3}\delta_{n_2,2}   \\ \delta_{n_1,3}\delta_{n_2,3}     
\end{pmatrix}.
\end{align}

Then, the $\hat{A}$ operator can be expressed as follows:
\begin{equation}
\scalebox{0.7}{$
\hat A = \theta\,
\left(
\begin{array}{cccccccccccccccc}
 0 & 0 & 0 & 0 & 0 & 0 & 0 & 0 & 0 & 0 & 0 & 0 & 0 & 0 & 0 & 0 \\
 0 & 0 & 0 & 0 & -1 & 0 & 0 & 0 & 0 & 0 & 0 & 0 & 0 & 0 & 0 & 0 \\
 0 & 0 & 0 & 0 & 0 & -\sqrt{2} & 0 & 0 & 0 & 0 & 0 & 0 & 0 & 0 & 0 & 0 \\
 0 & 0 & 0 & 0 & 0 & 0 & -\sqrt{3} & 0 & 0 & 0 & 0 & 0 & 0 & 0 & 0 & 0 \\
 0 & 1 & 0 & 0 & 0 & 0 & 0 & 0 & 0 & 0 & 0 & 0 & 0 & 0 & 0 & 0 \\
 0 & 0 & \sqrt{2} & 0 & 0 & 0 & 0 & 0 & -\sqrt{2} & 0 & 0 & 0 & 0 & 0 & 0 & 0 \\
 0 & 0 & 0 & \sqrt{3} & 0 & 0 & 0 & 0 & 0 & -2 & 0 & 0 & 0 & 0 & 0 & 0 \\
 0 & 0 & 0 & 0 & 0 & 0 & 0 & 0 & 0 & 0 & -\sqrt{6} & 0 & 0 & 0 & 0 & 0 \\
 0 & 0 & 0 & 0 & 0 & \sqrt{2} & 0 & 0 & 0 & 0 & 0 & 0 & 0 & 0 & 0 & 0 \\
 0 & 0 & 0 & 0 & 0 & 0 & 2 & 0 & 0 & 0 & 0 & 0 & -\sqrt{3} & 0 & 0 & 0 \\
 0 & 0 & 0 & 0 & 0 & 0 & 0 & \sqrt{6} & 0 & 0 & 0 & 0 & 0 & -\sqrt{6} & 0 & 0 \\
 0 & 0 & 0 & 0 & 0 & 0 & 0 & 0 & 0 & 0 & 0 & 0 & 0 & 0 & -3 & 0 \\
 0 & 0 & 0 & 0 & 0 & 0 & 0 & 0 & 0 & \sqrt{3} & 0 & 0 & 0 & 0 & 0 & 0 \\
 0 & 0 & 0 & 0 & 0 & 0 & 0 & 0 & 0 & 0 & \sqrt{6} & 0 & 0 & 0 & 0 & 0 \\
 0 & 0 & 0 & 0 & 0 & 0 & 0 & 0 & 0 & 0 & 0 & 3 & 0 & 0 & 0 & 0 \\
 0 & 0 & 0 & 0 & 0 & 0 & 0 & 0 & 0 & 0 & 0 & 0 & 0 & 0 & 0 & 0 \\
\end{array}
\right).
$}
\end{equation}

From here:
\begin{equation}
\scalebox{0.5}{$
    \hat{U}_{BS}(\theta,0)=\left(\begin{array}{cccccccccccccccc}1 & 0 & 0 & 0 & 0 & 0 & 0 & 0 & 0 & 0 & 0 & 0 & 0 & 0 & 0 & 0\\
    0 & \cos{\left(\theta \right)} & 0 & 0 & - \sin{\left(\theta \right)} & 0 & 0 & 0 & 0 & 0 & 0 & 0 & 0 & 0 & 0 & 0\\
    0 & 0 & \cos^2{\left(\theta \right)} & 0 & 0 & - \frac{\sin{\left(2 \theta \right)}}{\sqrt{2}} & 0 & 0 & \sin^2(\theta) & 0 & 0 & 0 & 0 & 0 & 0 & 0\\
    0 & 0 & 0 & \cos^3\theta & 0 & 0 & -\sqrt{3}\sin\theta\cos^2\theta & 0 & 0 & \sqrt{3}\cos\theta\sin^2\theta& 0 & 0 & -\sin^3\theta& 0 & 0 & 0\\
    0 & \sin{\left(\theta \right)} & 0 & 0 & \cos{\left(\theta \right)} & 0 & 0 & 0 & 0 & 0 & 0 & 0 & 0 & 0 & 0 & 0\\
    0 & 0 & \frac{\sin{\left(2 \theta \right)}}{\sqrt{2}} & 0 & 0 & \cos{\left(2 \theta \right)} & 0 & 0 & - \frac{\sin{\left(2 \theta \right)}}{\sqrt{2}} & 0 & 0 & 0 & 0 & 0 & 0 & 0\\
    0 & 0 & 0 & \sqrt{3}\sin\theta\cos^2\theta & 0 & 0 & \frac{\cos{\left(\theta \right)}+3 \cos{\left(3 \theta \right)}}{4} & 0 & 0 & \frac{\sin{\left(\theta \right)}-3 \sin{\left(3 \theta \right)}}{4} & 0 & 0 & \sqrt{3}\cos\theta\sin^2\theta & 0 & 0 & 0\\
    0 & 0 & 0 & 0 & 0 & 0 & 0 & \frac{1+\cos{\left(2 \sqrt{3} \theta \right)}}{2} & 0 & 0 & - \frac{\sqrt{2} \sin{\left(2 \sqrt{3} \theta \right)}}{2} & 0 & 0 & \sin^2\left(\sqrt{3} \theta \right) & 0 & 0\\
    0 & 0 & \sin^2(\theta) & 0 & 0 & \frac{\sin{\left(2 \theta \right)}}{\sqrt{2}} & 0 & 0 & \cos^2(\theta) & 0 & 0 & 0 & 0 & 0 & 0 & 0\\
    0 & 0 & 0 & \sqrt{3}\cos\theta\sin^2\theta & 0 & 0 & \frac{-\sin{\left(\theta \right)}+3 \sin{\left(3 \theta \right)}}{4} & 0 & 0 & \frac{\cos{\left(\theta \right)}+3 \cos{\left(3 \theta \right)}}{4} & 0 & 0 & - \sqrt{3} \sin\theta\cos^2\theta & 0 & 0 & 0\\
    0 & 0 & 0 & 0 & 0 & 0 & 0 & \frac{\sin{\left(2 \sqrt{3} \theta \right)}}{\sqrt{2}} & 0 & 0 & \cos{\left(2 \sqrt{3} \theta \right)} & 0 & 0 & - \frac{\sin{\left(2 \sqrt{3} \theta \right)}}{\sqrt{2}} & 0 & 0\\
    0 & 0 & 0 & 0 & 0 & 0 & 0 & 0 & 0 & 0 & 0 & \cos{\left(3 \theta \right)} & 0 & 0 & - \sin{\left(3 \theta \right)} & 0\\
    0 & 0 & 0 & \sin^3\theta & 0 & 0 & \sqrt{3}\cos\theta\sin^2\theta & 0 & 0 & \sqrt{3} \sin\theta\cos^2\theta & 0 & 0 & \cos^3\theta & 0 & 0 & 0\\
    0 & 0 & 0 & 0 & 0 & 0 & 0 & \sin^2\left(\sqrt{3} \theta \right) & 0 & 0 & \frac{\sin{\left(2 \sqrt{3} \theta \right)}}{\sqrt{2}} & 0 & 0 & \cos^2\left(\sqrt{3} \theta \right)& 0 & 0\\
    0 & 0 & 0 & 0 & 0 & 0 & 0 & 0 & 0 & 0 & 0 & \sin{\left(3 \theta \right)} & 0 & 0 & \cos{\left(3 \theta \right)} & 0\\
    0 & 0 & 0 & 0 & 0 & 0 & 0 & 0 & 0 & 0 & 0 & 0 & 0 & 0 & 0 & 1\end{array}\right)
$}
\end{equation}

The transformation $\hat{U}_{BS} = e^{\hat{A}}$ can be 
decomposed according to the total number of photons involved,
i.e., $n=n_1+n_2$. Then the transformation of the basis states 
can be written as the following direct sum:
\begin{equation}
U^{N=4}_{BS}(\theta,0) = \bigoplus_{i=0}^6 U_i.
\end{equation}

We emphasize that the blocks obtained below are the exact blocks of the beam-splitter unitary generated within the truncated Hilbert space. They coincide with the corresponding infinite-dimensional optical transformation for total photon-number sectors \(n\leq N-1\); for \(n\geq N\), they define the unitary truncated-space dynamics but need not coincide with the projection of the full optical evolution.

We find: 
\begin{itemize}
\item $n=0$ (no photons) - basis ($\ket{0,0}$)
\[
U_{0} \;=\;
\begin{pmatrix}
1
\end{pmatrix}.
\]
\item $n=1$ (1 photon) - basis ($\ket{0,1},\ket{1,0}$)
\[
U_{1} \;=\;
\begin{pmatrix}
\cos\theta & -\sin\theta\\
\sin\theta & \cos\theta
\end{pmatrix}.
\]
\item $n=2$ (2 photons) - basis ($\ket{0,2},\ket{1,1},\ket{2,0}$)
\[
U_{2} \;=\;
\begin{pmatrix}
\cos^2\theta & -\dfrac{\sin 2\theta}{\sqrt2} & \sin^2\theta\\
\dfrac{\sin 2\theta}{\sqrt2} & \cos 2\theta & -\dfrac{\sin 2\theta}{\sqrt2}\\
\sin^2\theta & \dfrac{\sin 2\theta}{\sqrt2} & \cos^2\theta
\end{pmatrix}.
\]

Worth noticing is that for the symmetric case $\theta=\pi/4$,
the interference properties leading to the Hong–Ou–Mandel effect
are recovered. Namely, one can observe that $U_{2}(\pi/4,0)\ket{1,1}=
\frac{1}{\sqrt{2}}(\ket{0,2}-\ket{2,0})$.

\item $n=3$ (3 photons)  - basis ($\ket{0,3},\ket{1,2},\ket{2,1},\ket{3,0}$)
\[
U_{3} =\begin{pmatrix}
\cos^3\theta & -\sqrt3\,\sin\theta\,\cos^2\theta & \sqrt3\,\sin^2\theta \cos\theta & -\sin^3\theta\\
\sqrt3\,\sin\theta\,\cos^2\theta & (3\cos^2\theta-2)\,\cos\theta & (3\sin^2\theta-2)\,\sin\theta & \sqrt3\,\sin^2\theta \cos\theta\\
\sqrt3\,\sin^2\theta \cos\theta & (2-3\sin^2\theta)\,\sin\theta & (3\cos^2\theta-2)\,\cos\theta & -\sqrt3\,\sin\theta\,\cos^2\theta\\
\sin^3\theta & \sqrt3\,\sin^2\theta \cos\theta & \sqrt3\,\sin\theta\,\cos^2\theta & \cos^3\theta
\end{pmatrix}.
\]
\item $n=4$ (4 photons)  - basis ($\ket{1,3},\ket{2,2},\ket{3,1}$)
\[
U_{4} = \begin{pmatrix}
\cos^2(\sqrt3\theta) & -\dfrac{\sin (2\sqrt3\theta)}{\sqrt2} & \sin^2(\sqrt3\theta)\\
\dfrac{\sin (2\sqrt3\theta)}{\sqrt2} & \cos (2\sqrt3\theta) & -\dfrac{\sin (2\sqrt3\theta)}{\sqrt2}\\
\sin^2(\sqrt3\theta) & \dfrac{\sin (2\sqrt3\theta)}{\sqrt2} & \cos^2(\sqrt3\theta)
\end{pmatrix}.
\]
\item $n=5$ (5 photons) - basis ($\ket{2,3},\ket{3,2}$)
\[
U_{5} =\begin{pmatrix}
\cos3\theta & -\sin3\theta\\
\sin3\theta & \cos3\theta
\end{pmatrix}.
\]
\item $n=6$ (6 photons) - basis ($\ket{3,3}$)
\[
U_{6} = \begin{pmatrix}
1
\end{pmatrix}.
\]
\end{itemize}

Please notice that the dimensions of the contributing space decrease 
starting from the occupation number $n=4$, which guarantees unitarity 
of the truncation. This is different from the untruncated case where 
the dimension grows linearly with the photon occupation number, 
following the $SU(2)$ element decomposition into the irreducible 
representations (via Wigner matrices). Worth emphasizing is that in 
the truncated framework, relations between the $n=1$ and $n=5$ 
($\theta \rightarrow 3 \theta$) as well $n=2$ and $n=4$ ($\theta \rightarrow \sqrt{3} \theta$) hold, which will later simplify 
construction of the corresponding quantum circuits. 

\subsubsection{Quantum circuit for the beam splitter}

In this approach, we implement, block by block, the actions of the 
$U_{i}$ matrices on a given number of photons. The total 
quantum circuit is a joint action of the contributions presented 
below. Importantly, the order of the contributions is irrelevant, 
since the blocks with a given number of photons commute. According 
to the number of photons involved, the contributions to the total 
quantum circuit for the $N=4$ truncation of the beam splitter are: 

\begin{itemize}
\item \textbf{n=0} - We do nothing.

\item \textbf{n=1} - $U_1$ can be implemented as a single qubit gate. It acts in subspace $\{\ket{0001},\ket{0100}\}$, we need to rotate this base to one where basis vectors differ only at one qubit, for example, $\{\ket{0001},\ket{0101}\}$, which can be achieved using $CX_{1\rightarrow3}$. Then, we apply a single-qubit gate $R_y(2\theta)$ to the second qubit controlled by all the remaining qubits to ensure that it acts only in this two-dimensional subspace, which leads to the following quantum circuit: 

\begin{center}
$U_1$
$\equiv$
\begin{quantikz}[baseline=(current bounding box.center)]
&\qw&\octrl{1}&\qw&\qw\\
&\ctrl{2}&\gate{R_y(2\theta)}&\ctrl{2}&\qw\\
&\qw&\octrl{-1}&\qw&\qw\\
&\targ{}&\ctrl{-1}&\targ{}&\qw\\
\end{quantikz}
\end{center}

\item  \textbf{n=2} - $U_2$ can be implemented using an extended
two-qubit gate:
$$
\tilde{U}_{2} \;=\;
\begin{pmatrix}
1&0&0&0\\
0&\cos^2\theta & -\dfrac{\sin 2\theta}{\sqrt2} & \sin^2\theta\\
0&\dfrac{\sin 2\theta}{\sqrt2} & \cos 2\theta & -\dfrac{\sin 2\theta}{\sqrt2}\\
0&\sin^2\theta & \dfrac{\sin 2\theta}{\sqrt2} & \cos^2\theta
\end{pmatrix}.
$$

Like in the case of $U_1$  we need to rotate the base to one 
where basis vectors differ only at one qubit pairwise. For example, from $\{\ket{0010}, \ket{0101}, \ket{1000}\}$ to $\{\ket{1101}, \ket{1110}, \ket{1111}\}$ using $X_2X_3CX_{3\rightarrow0}CX_{2\rightarrow0}CX_{\bar3\rightarrow1}$. Then we apply a two-qubit gate $\tilde{U}_2$ to the third and fourth qubits controlled by all remaining qubits to ensure that it acts only in this four-dimensional subspace. Gate $\tilde{U}_2$ can be decomposed as follows:
\begin{center}
    $\tilde{U}_2$
    $\equiv$
    \begin{quantikz}
    \lstick{$q_0$} & \gategroup[2,steps=3,style={dashed}]{$W^\dagger$} &
      \gate{R_y(\text{-}\frac{\pi}{2})} & \ctrl{1} &
      \gate{R_y(2\theta)} & \ctrl{1} \gategroup[2,steps=3,style={dashed}]{$W$}&
      \gate{R_y(\frac{\pi}{2})} & & \\
    \lstick{$q_1$} & \gate{X} & \ctrl{-1} & \targ{} &
      \gate{R_y(2\theta)} & \targ{} & \ctrl{-1} & \gate{X} &
    \end{quantikz}
\end{center}

So, the quantum circuit for the full $U_2$ block is:
\begin{center}
    $U_2$
    $\equiv$
    \begin{quantikz}
     & \targ{} & \targ{} & & & & \ctrl{1} & \ctrl{1} & & & & \targ{} & \targ{} & \\
     & & & \targ{} & & & \ctrl{1} & \ctrl{2} & & & \targ{} & & & \\
     & & \ctrl{-2} & & \gate{X} & \gate[2]{W^\dagger} & \gate{R_y(2\theta)} & & \gate[2]{W} & \gate{X} & & \ctrl{-2} & & \\
     & \ctrl{-3} & & \octrl{-2} & \gate{X} & & & \gate{R_y(2\theta)} & & \gate{X} & \octrl{-2} & & \ctrl{-3} & 
    \end{quantikz}
\end{center}

\item  \textbf{n=3} - Similarly as before, we change the basis from $\{\ket{1100},\ket{1001},\ket{0110},\ket{0011}\}$ to \newline
$\{\ket{1100},\ket{1101},\ket{1110},\ket{1111}\}$ to perform 
the action on the corresponding 2-qubit subspace. The 
2-qubit action is given by: 
\begin{center}
    $\tilde{U}_3$
    $\equiv$
    \begin{quantikz}
    \lstick{$q_0$} & \gate{S}\gategroup[2,steps=3,style={dashed}]{$M$} & &
      \targ{} &
      \gate{R_y(2\theta)} & \targ{} \gategroup[2,steps=3,style={dashed}]{$M^\dagger$}& & \gate{S^\dagger}& \\
    \lstick{$q_1$} & \gate{S} & \gate{H} & \ctrl{-1} &\gate{B} & \ctrl{-1} & \gate{H} & \gate{S^\dagger} &
    \end{quantikz},
\end{center}
where for the further utility we introduced the operation $M$.
Then, action of the full block $U_3$, inolving the basis change 
and the controlled $\tilde{U}_3$ part reads: 

\begin{center}
    $U_3$
    $\equiv$
    \begin{quantikz}
     & \targ{} & & & & \ctrl{1} & \ctrl{1} & \ctrl{1} & \ctrl{1} & & & & \targ{} & \\
     & & \targ{} & & & \ctrl{1} & \ctrl{2} & \ctrl{2} & \ctrl{2} & & & \targ{} & & \\
     & \ctrl{-2} & & \gate{X} & \gate[2]{M} & \gate{R_y(2\theta)} & & & & \gate[2]{M^\dagger} & \gate{X} & & \ctrl{-2} & \\
     & & \ctrl{-2} & \gate{X} & & & \gate{R_z(\alpha)} & \gate{R_y(\beta)} & \gate{R_z(\alpha)} & & \gate{X} & \ctrl{-2} & & 
    \end{quantikz}
\end{center}

The $B$ gate has the following decomposition:
\begin{equation}
B=\begin{pmatrix}
    \cos(2\theta)+i\frac{\sqrt{3}}{2}\sin(2\theta)& -\frac{\sin(2\theta)}{2}\\
    \frac{\sin(2\theta)}{2} & \cos(2\theta)-i\frac{\sqrt{3}}{2}\sin(2\theta)\\
\end{pmatrix}
=R_z(\alpha)R_y(\beta)R_z(\alpha),
\end{equation}
where the angles involved can be expressed as follows:
\begin{align}
\alpha(\theta) &= -\operatorname{atan2}\big(\sqrt3\sin 2\theta,\; 2\cos 2\theta\big), \\ 
\qquad  
\beta(\theta) &= 2\arcsin\Big(\tfrac{1}{2}\sin 2\theta\Big).
\end{align}

\item \textbf{n=4} - Here, we change base $\{\ket{0111},\ket{1010},\ket{1101}\}$ to $\{\ket{0000},\ket{0001},\ket{0010}\}$ using three 
CNOT gates. Then the 3-level action of $U_4$ can 
be associated with the $2$-qubit operation $\tilde{U}_4$:

$$
\tilde{U}_4=
\begin{pmatrix}
1&0&0&0\\
0&\cos^2(\sqrt3\theta) & -\dfrac{\sin (2\sqrt3\theta)}{\sqrt2} & \sin^2(\sqrt3\theta)\\
0&\dfrac{\sin (2\sqrt3\theta)}{\sqrt2} & \cos (2\sqrt3\theta) & -\dfrac{\sin (2\sqrt3\theta)}{\sqrt2}\\
0&\sin^2(\sqrt3\theta) & \dfrac{\sin (2\sqrt3\theta)}{\sqrt2} & \cos^2(\sqrt3\theta)
\end{pmatrix}
$$
which is the same as $\tilde{U}_2$ but with $\theta\to\sqrt3\,\theta$. The quantum circuit 
corresponding to the matrix $\tilde{U}_4$ can be 
expressed as: 
\begin{center}
    $\tilde{U}_4$
    $\equiv$
    \begin{quantikz}
    \lstick{$q_0$} & \gategroup[2,steps=3,style={dashed}]{$W^\dagger$} &
      \gate{R_Y(-\frac{\pi}{2})} & \ctrl{1} &
      \gate{R_y(2\sqrt{3}\theta)} & \ctrl{1} \gategroup[2,steps=3,style={dashed}]{$W$}&
      \gate{R_y(\frac{\pi}{2})} & & \\
    \lstick{$q_1$} & \gate{X} & \ctrl{-1} & \targ{} &
      \gate{R_y(2\sqrt{3}\theta)} & \targ{} & \ctrl{-1} & \gate{X} &
    \end{quantikz}
\end{center}

From here, the total circuit for $U_4$, involving the 
basis change and the controlled $\tilde{U}_4$ operation, 
reads: 
\begin{center}
    $U_4$
    $\equiv$
    \begin{quantikz}
     & \targ{} & \targ{} & & & \octrl{1} & \octrl{1} & & & \targ{} & \targ{} & \\
     & & & \targ{} & & \octrl{1} & \octrl{2} & & \targ{} & & & \\
     & & \ctrl{-2} & & \gate[2]{W^\dagger} & \gate{R_y(-2\sqrt{3}\theta)} & & \gate[2]{W} & & \ctrl{-2} & & \\
     & \ctrl{-3} & & \ctrl{-2} & & & \gate{R_y(-2\sqrt{3}\theta)} & & \ctrl{-2} & & \ctrl{-3} & 
    \end{quantikz}
\end{center}

\item \textbf{n=5} - In this case, similarly to the
$U_1$ case, we change the basis $\{\ket{1011},\ket{1110}\}$ to $\{\ket{1011},\ket{1111}\}$ using a single CNOT gate. 
Then the corresponding 1-qubit rotation can be applied. 
Controlling of the rotation is required to constrain action
of the rotation operator to the desired $n=5$ subspace.
The resulting quantum circuit is: 

\begin{center}
$U_5$
$\equiv$
\begin{quantikz}[baseline=(current bounding box.center)]
&\qw&\ctrl{1}&\qw&\qw\\
&\ctrl{2}&\gate{R_y(6\theta)}&\ctrl{2}&\qw\\
&\qw&\ctrl{-1}&\qw&\qw\\
&\targ{}&\ctrl{-1}&\targ{}&\qw\\
\end{quantikz}
\end{center}

\item \textbf{n=6} - We do nothing.

\end{itemize}

\subsubsection{Reduced quantum circuit for the beam splitter}
\label{sec:bs-reduced}

In the block-by-block construction, every $U_n$ needs its own CNOT basis 
change and rotations with two or three controls. After compilation to CNOT 
and single-qubit gates this gives 115 CNOTs. Here, we provide an optimised
circuit which involves 28 two-qubit gates. Several blocks share one basis 
change, and the block structure is carried by the basis change instead of 
by additional controls. The optimisation process has been conducted with 
the use of Claude AI (Opus 5). 

The optimized circuit takes the following form: 
\begin{center}
$\hat{U}_{BS}(\theta,0)$
$\equiv$
\begin{quantikz}[column sep=0.25cm]
\lstick{$q_0$} & \targ{} & & & \gate[4]{V_{\mathrm{odd}}} & & \targ{} & & \gate[4]{V_{\mathrm{even}}} & & \targ{} & & \targ{} & \\
\lstick{$q_1$} & & \targ{} & & & & & & & & & \targ{} & & \\
\lstick{$q_2$} & \ctrl{-2} & & \gate[2]{M} & & \gate[2]{M^\dagger} & & \gate[2]{M} & & \gate[2]{M^\dagger} & & & \ctrl{-2} & \\
\lstick{$q_3$} & & \ctrl{-2} & & & & \ctrl{-3} & & & & \ctrl{-3} & \ctrl{-2} & &
\end{quantikz}
\end{center}

Here, the $M$ gate corresponds to the block introduced for the $U_{3}$ circuit. 
Furthermore, $V_{\mathrm{odd}}$ implements $U_1$, $U_3$ and $U_5$ sectors 
and acts only for $q_1=1$:
\begin{center}
$V_{\mathrm{odd}}\equiv$
\resizebox{0.9\linewidth}{!}{%
\begin{quantikz}[column sep=0.2cm]
\lstick{$q_0$} & & \ctrl{2} & \ctrl{3} & & & & & \ctrl{2} & \ctrl{3} & & \\
\lstick{$q_1$} & & & & & \ctrl{1} & \ctrl{2} & & & & & \\
\lstick{$q_2$} & \gate{R_x(-\frac{\pi}{4})} & \gate{Y} & & \gate{R_x(-\frac{\pi}{4})} & \gate{R_y(-2\theta)} & & \gate{R_x(\frac{\pi}{4})} & \gate{Y} & & \gate{R_x(\frac{\pi}{4})} & \\
\lstick{$q_3$} & \gate{R_x(\frac{5\pi}{12})} & & \gate{Y} & \gate{R_x(\frac{\pi}{12})} & & \gate{R_y(-4\theta)} & \gate{R_x(-\frac{\pi}{12})} & & \gate{Y} & \gate{R_x(-\frac{5\pi}{12})} &
\end{quantikz}%
}
\end{center}

Similarly, $V_{\mathrm{even}}$ implements $U_2$ and $U_4$ sectors 
and acts only for $q_1=0$:
\begin{center}
$V_{\mathrm{even}}\equiv$
\resizebox{0.9\linewidth}{!}{%
\begin{quantikz}[column sep=0.15cm]
\lstick{$q_0$} & \ctrl{2} & & & & & \ctrl{2} & & & & \ctrl{2} & & & & \ctrl{3} & & & & \ctrl{3} & & \ctrl{2} & \\
\lstick{$q_1$} & & & & \ctrl{1} & & & & \ctrl{1} & & & & \ctrl{2} & & & & \ctrl{2} & & & & & \\
\lstick{$q_2$} & \gate{Y} & \gate{R_x(-\frac{\pi}{4})} & \gate{R_y(\xi_+)} & \targ{} & \gate{R_y(\xi_+)} & \targ{} & \gate{R_y(\xi_-)} & \targ{} & \gate{R_y(\xi_-)} & \targ{} & & & & & & & & & \gate{R_x(\frac{\pi}{4})} & \gate{Y} & \\
\lstick{$q_3$} & & \gate{R_x(\frac{\pi}{4})} & & & & & & & & & \gate{R_y(\xi_+)} & \targ{} & \gate{R_y(\xi_+)} & \targ{} & \gate{R_y(\xi_-)} & \targ{} & \gate{R_y(\xi_-)} & \targ{} & \gate{R_x(-\frac{\pi}{4})} & &
\end{quantikz}%
}
\end{center}
where:
\begin{equation}
\xi_\pm=-\frac{\sqrt3\pm1}{2}\,\theta .
\end{equation}

In total, the circuit involves 22 CNOT and 6 controlled-$Y$ gates, 
36 single-qubit gates, and the depth is 45. For comparison, compiling 
to CNOT and single-qubit gates with Qiskit (optimization level 3) 
gives 115 CNOTs for the block-by-block construction and 93 CNOTs for 
a generic quantum Shannon decomposition of the same $16\times16$ unitary 
\cite{Shende2006}. All $\theta$-dependent angles are linear in $\theta$. 
In particular, the $\operatorname{atan2}$-dependent gate $B$ of $U_3$ no 
longer appears.

\section{Benchmarking}

In this section, we assess the accuracy of the proposed qubit-based representation of truncated quantum optical operations. The benchmarking analysis
is aimed at quantifying the precision with which photonic states and optical transformations are reproduced within a finite-dimensional Hilbert space, as well
as at identifying the regime of validity of the adopted truncation scheme.

Let us first observe that the basic operators discussed in this work can be divided into two groups: \textit{state-generating Gaussian operations} (displacement operator and squeezing operator) and \textit{number-preserving interferometric operations} (phase shift, beam splitter). Moreover, in standard Gaussian Boson Sampling architectures, the state-generating Gaussian operations are applied only at the beginning of the circuit and therefore act on the initial vacuum state, whereas interferometric operations act on the subsequently generated quantum state.

Passive Gaussian operations preserve the total photon number. However, for a truncation imposed independently on each optical mode, photon-number preservation alone does not guarantee exact agreement with the corresponding infinite-dimensional optical transformation. Exact agreement is guaranteed whenever the complete fixed-total-photon-number sector involved in the evolution is contained within the truncated Hilbert space. In particular, for a per-mode cutoff \(N\), passive linear-optical transformations are represented exactly for states with total photon number \(n_{\mathrm{tot}}\leq N-1\). For higher photon-number sectors, the truncation may also modify the interferometric dynamics.

Consequently, in the low-photon-number regime relevant to the present implementation of Gaussian Boson Sampling, the dominant truncation errors originate from the state-generation stage, while the passive part of the circuit is exact within the above photon-number range. Below, we discuss the phase shift and beam splitter separately, showing that the former is exact throughout the truncated space, whereas for the latter exactness depends on whether the corresponding fixed-total-photon-number sector is fully contained within the truncation.

We therefore focus the quantitative benchmark primarily on the state-generating operators. The displacement and squeezing operators are benchmarked using photon-number-based measures that take into account the structure of the generated Gaussian states.

\subsection{Phase shift}
The phase shift operator depends exclusively on the photon number operator $\hat{N}$. In this work, truncation is performed directly in the Fock basis by restricting the total photon number, such that $\hat{N}$ is represented exactly within the $0,\dots,N-1$ photon sector. Since $\hat{N}$ is diagonal in the Fock basis, its action does not couple different photon-number sectors and therefore cannot induce population leakage or fidelity loss within the truncated space. Consequently, the phase shift operator is implemented exactly on all photon-number sectors supported by the truncation and does not introduce truncation-related errors. For this reason, no separate benchmarking of the phase shift operator is required.

\subsection{Coherent states} 

The mean photon number for a coherent state has a  simple form:
\begin{equation}
    \langle0|\hat D^{\dagger}(\alpha)\hat a^{\dagger}\hat a\hat D(\alpha)|0\rangle=|\alpha|^2=r^2,
\end{equation}
for $\alpha=re^{i\phi}$.
Denoting the truncated creation/annihilation operators and the corresponding truncated displacement operators as respectively $\hat b^{\dagger}_N,\hat b_N,\hat D_N(\alpha)$, we define the relative error of the mean photon number in the truncated coherent state as follows:
\begin{equation}
    \label{eq:relErrorD}
    d_R(\alpha,N)=\frac{| \langle0|\hat D_N^{\dagger}(\alpha)\hat b_N^{\dagger}\hat b_N\hat D_N(\alpha)|0\rangle- \langle0|\hat D^{\dagger}(\alpha)\hat a^{\dagger}\hat a\hat D(\alpha)|0\rangle|}{ \langle0|\hat D^{\dagger}(\alpha)\hat a^{\dagger}\hat a\hat D(\alpha)|0\rangle}=\frac{| \langle0|\hat D_N^{\dagger}(\alpha)\hat b_N^{\dagger}\hat b_N\hat D_N(\alpha)|0\rangle- r^2|}{r^2}.
\end{equation}

Such finite-dimensional analogues of coherent states, including states generated by truncated displacement-type constructions, have been studied previously in several formulations \cite{Buzek1992,Kuang1994,Miranowicz2001,Leonski2001}.

\begin{figure}[htbp]
    \centering
    \includegraphics[height=6.1 cm]{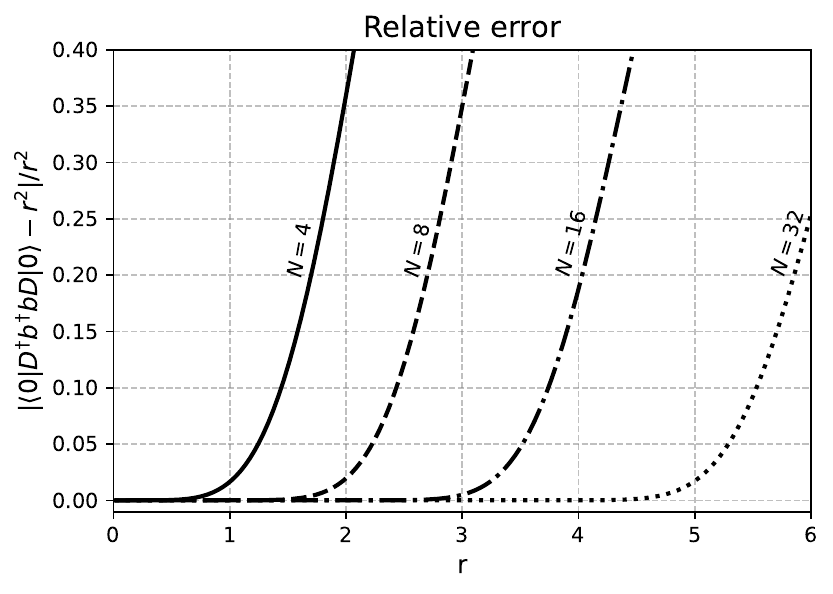}\hfill
    \includegraphics[height=6.1 cm]{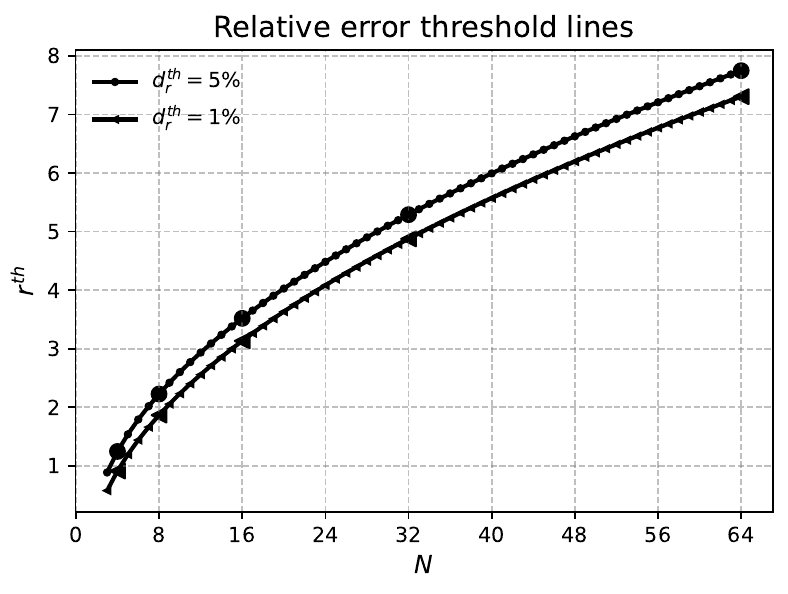}\hfill
    \caption{\textit{Left:} Relative error, $d_R(\alpha = r\cdot e^{i\phi}, N),\ \phi=0,\ N\in\{4,8,16,32\}$. \textit{Right:} The value of $r$ for which one obtains the relative error $d^{th}_r$ value of $5\%$ and $1\%$ as a function of the Fock space dimension, $N$. The scatter points corresponding to $N$ values achievable within the qubit-based architecture are enlarged.}
    \label{fig:dRDisplaced}
\end{figure}

The values of $d_R$ as a function of the displacement amplitude $r$ are shown 
in the left panel of Fig. \ref{fig:dRDisplaced}. The resulting maximal allowed 
values of $r$ for the errors  $1\%$ and  $5\%$ are shown in the right panel 
of the same figure. 

\subsection{Squeezed states}

In order to characterize the error of approximation of the squeezing operation, we utilize the fact that the mean photon number for a squeezed vacuum state has a  simple form:
\begin{equation}
    \langle0|\hat S^{\dagger}(z)\hat a^{\dagger}\hat a\hat S(z)|0\rangle=\sinh^2{r},
\end{equation}
for $z=re^{i\phi}$.
Let us denote the truncated creation/annihilation operators and the corresponding truncated squeezing operators as respectively $\hat b^{\dagger}_N,\hat b_N,\hat S_N(z)$. Then the relative error of the mean photon number is specified as:
\begin{equation}
    \label{eq:relErrorSq}
    d_R(z,N)=\frac{| \langle0|\hat S_N^{\dagger}(z)\hat b_N^{\dagger}\hat b_N\hat S_N(z)|0\rangle- \langle0|\hat S^{\dagger}(z)\hat a^{\dagger}\hat a\hat S(z)|0\rangle|}{ \langle0|\hat S^{\dagger}(z)\hat a^{\dagger}\hat a\hat S(z)|0\rangle}=\frac{| \langle0|\hat S_N^{\dagger}(z)\hat b_N^{\dagger}\hat b_N\hat S_N(z)|0\rangle- \sinh^2{r}|}{\sinh^2{r}}.
\end{equation}
Assuming $N=4$, taking the formulas \eqref{b4} and \eqref{sq4} one obtains a simple analytic expression:
\begin{equation}
    \langle0|\hat S_4^{\dagger}(z)\hat b_4^{\dagger}\hat b_4\hat S_4(z)|0\rangle=2\sin^2\left(\frac{r}{\sqrt 2}\right),
\end{equation}
which gives a closed form for the relative error:
\begin{equation}
    \label{eq:relErrorSq4}
d_R(z,4)=\frac{\left|2\sin^2\left(\frac{r}{\sqrt 2}\right)- \sinh^2{r}\right|}{\sinh^2{r}}.
\end{equation}

\begin{figure}[htbp]
    \centering
    \includegraphics[height=6 cm]{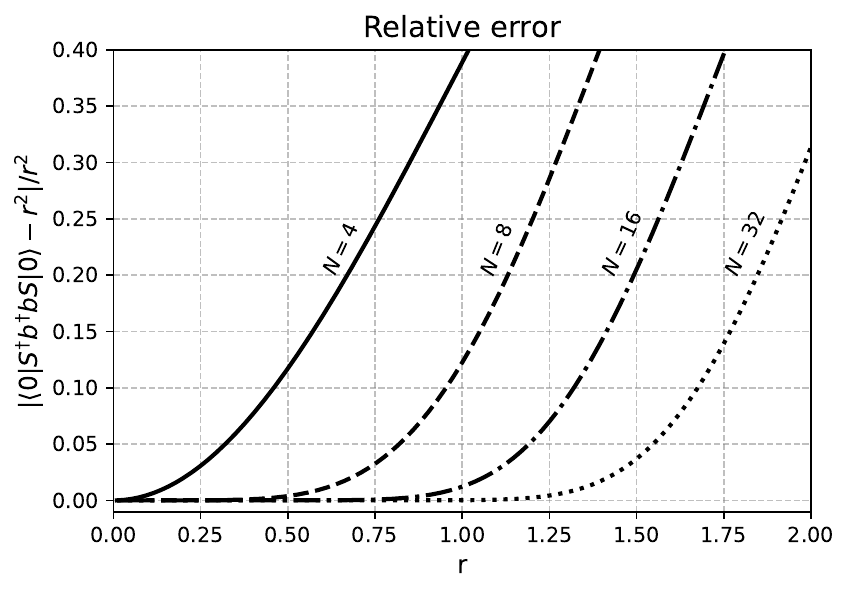}\hfill
    \includegraphics[height=6 cm]{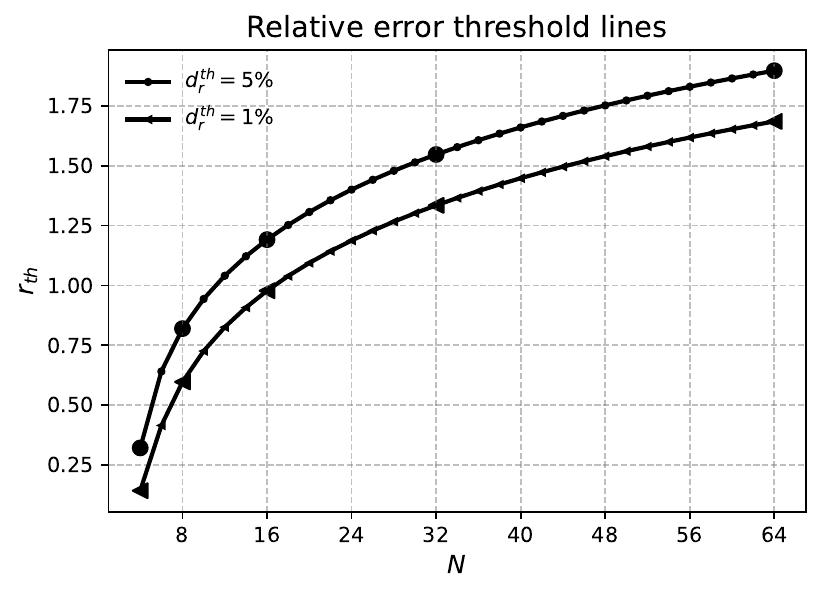}\hfill
    \caption{\textit{Left:} Relative error, $d_R(\alpha = r\cdot e^{i\phi}, N),\ \phi=0,\ N\in\{4,8,16,32\}$. \textit{Right:} The value of $r$ for which one obtains the relative error $d^{th}_r$ value of $5\%$ and $1\%$ as a function of the Fock space dimension, $N$. The scatter points corresponding to $N$ values achievable within the qubit-based architecture are enlarged.}
    \label{fig:dRSqueezed}
\end{figure}

The values of $d_R$ as a function of the displacement amplitude $r$ are shown 
in the left panel of Fig. \ref{fig:dRSqueezed}. The resulting maximal allowed 
values of $r$ for the errors  $1\%$ and  $5\%$ are shown in the right panel 
of the same figure. 

\subsection{Beam splitter}

Beam splitter is another passive circuit element which preserves the total number of photons. Remembering the commutation relations,
\begin{align}
    [\hat{N}, \hat{b}] = -\hat{b},\ [\hat{N},\hat{b}^{\dagger}]= \hat{b}^{\dagger},
\end{align}
and defining the total number of particles as $\hat{N}_1+\hat{N}_2$ we have:
\begin{align}
    [\hat{b}_1^{\dagger} \hat{b}_2 - \hat{b}_1 \hat{b}_2^{\dagger}, \hat{N_1}+\hat{N_2}] = 
    [\hat{b}_1^{\dagger} \hat{b}_2, \hat{N_1}+\hat{N_2}] - [\hat{b}_1 \hat{b}_2^{\dagger},\hat{N_1}+\hat{N_2}] = 0.
\end{align}

The beam splitter couples Fock states \(|n_1,n_2\rangle\) only within sectors of fixed total photon number \(n=n_1+n_2\). In the full Fock space, such a sector is spanned by
$$ \{|0,n\rangle,|1,n-1\rangle,\ldots,|n,0\rangle\}. $$
For the per-mode truncation \(|0\rangle,\ldots,|N-1\rangle\), this entire sector is contained in the truncated Hilbert space only for \(n\leq N-1\). In this regime, the truncated beam-splitter generator coincides with the restriction of the full generator and the resulting evolution is exact. For \(n\geq N\), some states belonging to the corresponding full photon-number sector lie outside the truncated space, and the unitary generated by the truncated operators generally differs from the projection of the infinite-dimensional beam-splitter evolution. Thus, photon-number conservation guarantees block-diagonality, but not exactness of the truncation in the higher-number sectors.

\section{Description of the compiler}

\begin{figure}[h]
\centering
\begin{tikzpicture}[node distance=0.5cm]

  \node[pipebox=orange!15] (bb) {Blackbird source\\[1pt]\texttt{.xbb} file};
  \node[pipebox=orange!10, below=of bb] (parse) {\texttt{bb.loads()}\\[1pt]Blackbird parser};
  \node[pipebox=blue!10, below=of parse] (ast) {Operation list\\[1pt]\texttt{[\{op, args, modes\}]}};
  \node[pipebox=green!12, below=of ast] (decomp) {Gate decomposition\\[1pt]\texttt{CVOperation} classes};
  \node[pipebox=blue!10, below=of decomp] (qs) {PennyLane\\[1pt]\texttt{QuantumScript}};
  \node[pipebox=violet!12, below=of qs] (export) {\texttt{qml.to\_openqasm()}\\[1pt]+ strip \texttt{gphase}};
  \node[pipebox=red!12, below=of export] (qasm) {OpenQASM 2.0\\[1pt]string output};

  \foreach \a/\b in {bb/parse, parse/ast, ast/decomp, decomp/qs, qs/export, export/qasm} {
    \draw[-{Stealth[length=6pt]}, thick] (\a) -- (\b);
  }

  \node[left=1.2cm of parse, font=\scriptsize\itshape, text width=2.5cm, align=right]
    {External library:\\blackbird};
  \node[left=1.2cm of decomp, font=\scriptsize\itshape, text width=2.5cm, align=right]
    {\texttt{operations.py}\\per-gate logic};
  \node[left=1.2cm of export, font=\scriptsize\itshape, text width=2.5cm, align=right]
    {External library:\\PennyLane};

\end{tikzpicture}
\caption{End-to-end compilation pipeline from Blackbird to OpenQASM.}
\label{fig:pipeline}
\end{figure}
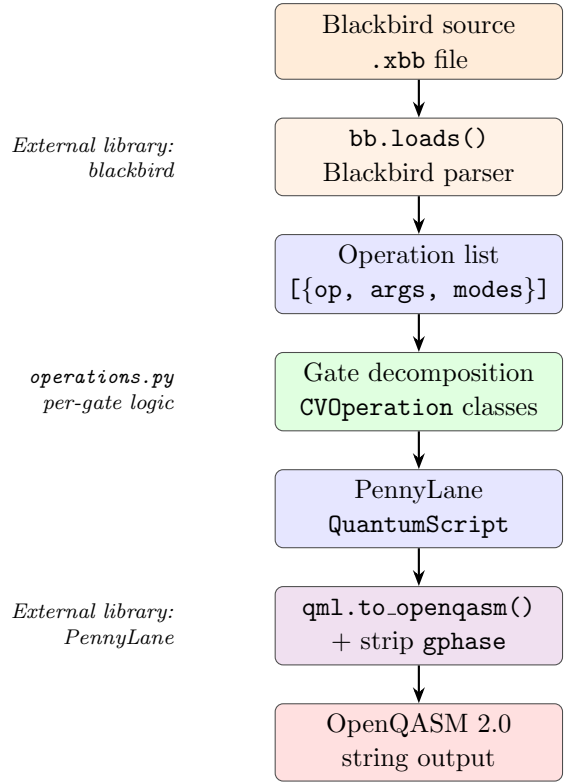

The CO$\Phi$ compiler \cite{Czelusta_CO_cophi_COmpiling} 
is organized into two modules: a high-level compiler driver 
(\texttt{compiler.py}) and a library of CV gate implementations
(\texttt{operations.py}). The implementation uses PennyLane for the 
construction and manipulation of qubit circuits and their subsequent 
export to OpenQASM \cite{Bergholm2018}. Fig.~\ref{fig:pipeline} shows 
the end-to-end compilation pipeline.

\subsection{Class Diagram}
\begin{sidewaysfigure}[h!]
    \centering
    \vspace{-2cm}\begin{tikzpicture}[node distance=1.0cm and 1.5cm]
    
      \node[abstractclass=3] (cvop) {
        \textbf{\textit{CVOperation}}
        \nodepart{second}
        \begin{tabular}{@{}l@{}}
          \texttt{fock\_cutoff: int}\\
          \texttt{num\_qubits\_per\_mode: int}
        \end{tabular}
        \nodepart{third}
        \begin{tabular}{@{}l@{}}
          \texttt{map\_mode\_to\_wires(mode)}\\
          \texttt{magic\_gate\_numerical\_matrix()}\\
          \texttt{magic\_gate\_decomposition(wires)}
        \end{tabular}
      };
    
      \node[umlclass, below left=3cm and 1.8cm of cvop] (phase) {
        \textbf{PhaseShift}
        \nodepart{second}
        \nodepart{third}
        \begin{tabular}{@{}l@{}}
          \texttt{symbolic\_matrix($\varphi$)}\\
          \texttt{numerical\_matrix($\varphi$)}\\
          \texttt{gate\_decomposition($\varphi$, mode)}
        \end{tabular}
      };
    
      \node[umlclass, right=0.4cm of phase] (disp) {
        \textbf{Displacement}
        \nodepart{second}
        \nodepart{third}
        \begin{tabular}{@{}l@{}}
          \texttt{symbolic\_matrix(r, $\varphi$)}\\
          \texttt{numerical\_matrix(r, $\varphi$)}\\
          \texttt{gate\_decomposition(r, $\varphi$, mode)}
        \end{tabular}
      };
    
      \node[umlclass, right=0.4cm of disp] (squeeze) {
        \textbf{Squeeze}
        \nodepart{second}
        \nodepart{third}
        \begin{tabular}{@{}l@{}}
          \texttt{symbolic\_matrix(r, $\varphi$)}\\
          \texttt{numerical\_matrix(r, $\varphi$)}\\
          \texttt{gate\_decomposition(r, $\varphi$, mode)}
        \end{tabular}
      };
    
      \node[umlclass, right=0.4cm of squeeze] (bs) {
        \textbf{BeamSplitter}
        \nodepart{second}
        \nodepart{third}
        \begin{tabular}{@{}l@{}}
          \texttt{symbolic\_matrix($\theta$, $\varphi$)}\\
          \texttt{numerical\_matrix($\theta$, $\varphi$)}\\
          \texttt{gate\_decomposition($\theta$, $\varphi$, modes)}
        \end{tabular}
      };
    
      \foreach \child in {phase, disp, squeeze, bs} {
        \draw[inherits] (\child.north) -- ++(0, 0.5) -| (cvop.south);
      }
    
      \node[umlclass, right=0.0cm and 4.5cm of cvop] (compiler) {
        \textbf{CompilerGBS}
        \nodepart{second}
        \begin{tabular}{@{}l@{}}
          \texttt{fock\_cutoff: int}\\
          \texttt{num\_qubits\_per\_mode: int}\\
          \texttt{\_SUPPORTED\_OPS: frozenset}\\
          \texttt{\_OP\_DISPATCH: dict}
        \end{tabular}
        \nodepart{third}
        \begin{tabular}{@{}l@{}}
          \texttt{compile\_to\_qasm(bb\_circuit)}\\
          \texttt{compile\_to\_pennylane(bb\_circuit)}\\
          \texttt{required\_qubits\_num(bb\_circuit)}\\
          \texttt{binary\_to\_photon\_meas(samples)}\\
          \texttt{map\_mode\_to\_wires(mode)}
        \end{tabular}
      };
    
      \draw[uses] (compiler.west) -- ++(-0.8,0) |- node[above, near start, font=\scriptsize\itshape]{uses} (cvop.east);
    
      \node[rectangle, draw, fill=gray!10, rounded corners=3pt,
            font=\small, align=center, minimum width=4cm, minimum height=1.2cm,
            below=3.5cm of disp] (utils) {
        \textbf{Utility functions}\\[2pt]
        \texttt{matrix\_to\_U3(A, wire)}\\
        \texttt{decompose\_kronecker(AB)}
      };
    
      \draw[uses] (disp.south) -- node[right, near end, font=\scriptsize\itshape]{uses} (utils.north);
    
    \end{tikzpicture}
    \caption{Class diagram of the GBS Compiler package.  Italic name denotes the
    abstract base class.  Dashed arrows indicate usage dependencies.}
    \label{fig:classes}
    \end{sidewaysfigure}
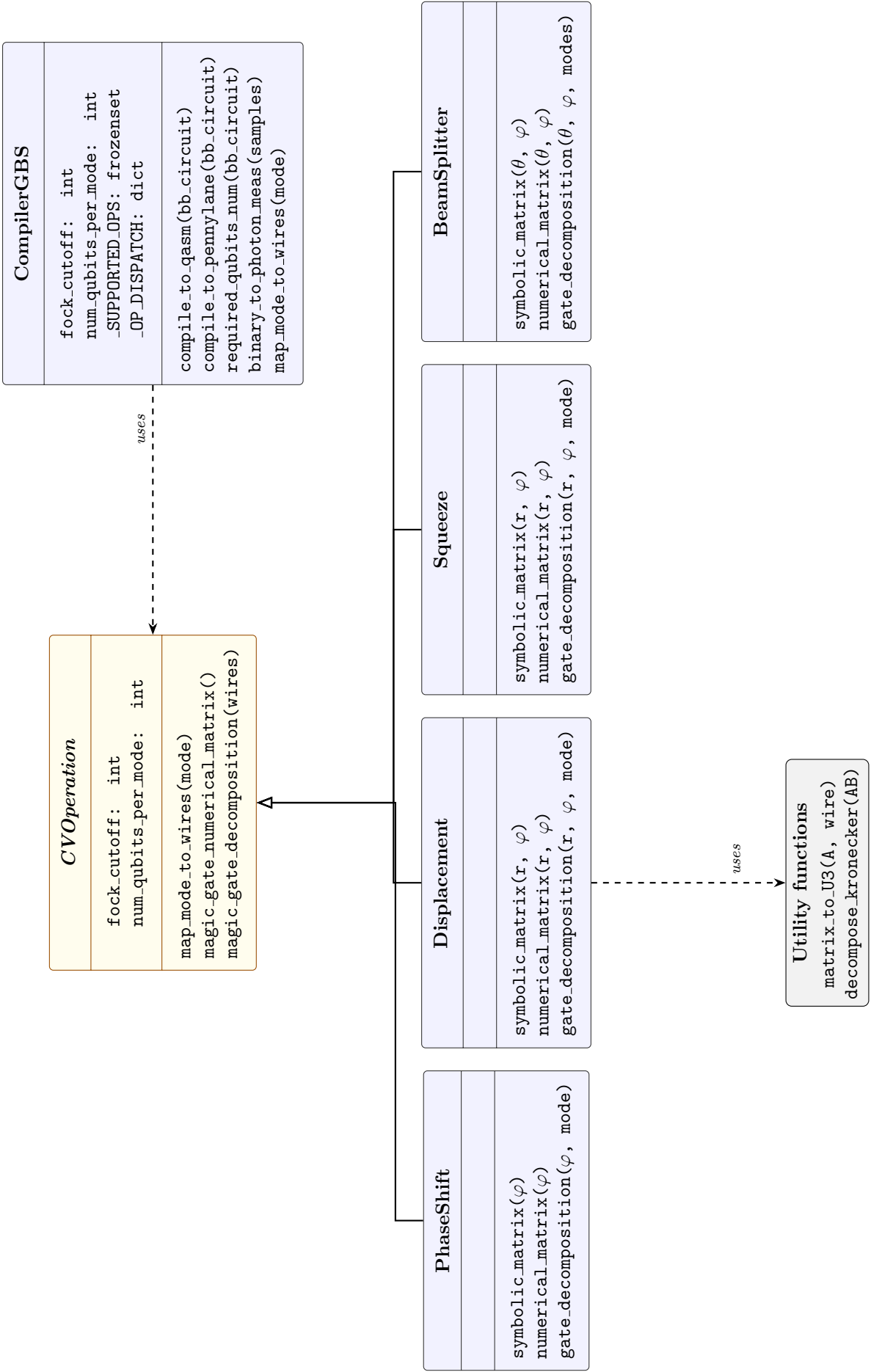

Fig.~\ref{fig:classes} shows the full class hierarchy.  \texttt{CVOperation} is the abstract base providing wire mapping and magic-basis utilities.  Four concrete subclasses implement the supported CV gates.  \texttt{CompilerGBS} consumes these classes through its dispatch table.

\subsection{Compilation Sequence}

\begin{figure}[h!]
\centering
\hspace*{-2cm}\begin{tikzpicture}[node distance=0.1cm]

  \node[actor] (user) {User};
  \node[actor, right=1cm of user] (comp) {CompilerGBS};
  \node[actor, right=1.0cm of comp] (bblib) {blackbird};
  \node[actor, right=1.0cm of bblib] (ops) {CVOperation};
  \node[actor, right=1.0cm of ops] (pl) {PennyLane};

  \foreach \n in {user, comp, bblib, ops, pl} {
    \draw[thick, gray!50] (\n.south) -- ++(0, -12.5);
  }

  \def\ystart{-1.2}
  \def\ystep{1.2}

  \draw[seqarrow] ([yshift=\ystart cm]user.south) --
    node[above, font=\scriptsize] {\texttt{compile\_to\_qasm(bb\_str)}}
    ([yshift=\ystart cm]comp.south);

  \pgfmathsetmacro{\y}{\ystart - \ystep}
  \draw[seqarrow] ([yshift=\y cm]comp.south) --
    node[above, font=\scriptsize] {\texttt{bb.loads(bb\_str)}}
    ([yshift=\y cm]bblib.south);

  \pgfmathsetmacro{\y}{\ystart - 2*\ystep}
  \draw[seqreturn] ([yshift=\y cm]bblib.south) --
    node[above, font=\scriptsize] {operations list}
    ([yshift=\y cm]comp.south);

  \pgfmathsetmacro{\y}{\ystart - 3*\ystep}
  \node[draw, dashed, rounded corners, fill=blue!3,
        minimum width=10.0cm, minimum height=2.8cm,
        anchor=north west, font=\scriptsize]
    at ([xshift=-0.3cm, yshift={\y cm + 0.3cm}]comp.south) {};
  \node[fill=blue!15, draw, font=\scriptsize\bfseries, rounded corners=2pt]
    at ([xshift=1.6cm, yshift={\y cm + 0.3cm}]comp.south) {loop: for each CV operation};

  \pgfmathsetmacro{\yi}{\ystart - 3.5*\ystep}
  \draw[seqarrow] ([yshift=\yi cm]comp.south) --
    node[above, font=\scriptsize] {\texttt{gate\_decomposition(params, mode)}}
    ([yshift=\yi cm]ops.south);

  \pgfmathsetmacro{\yi}{\ystart - 4.5*\ystep}
  \draw[seqreturn] ([yshift=\yi cm]ops.south) --
    node[above, font=\scriptsize] {\texttt{QuantumScript} (gates)}
    ([yshift=\yi cm]comp.south);

  \pgfmathsetmacro{\y}{\ystart - 6*\ystep}
  \draw[seqarrow] ([yshift=\y cm]comp.south) --
    node[above, font=\scriptsize] {\texttt{qml.to\_openqasm(script)}}
    ([yshift=\y cm]pl.south);

  \pgfmathsetmacro{\y}{\ystart - 7*\ystep}
  \draw[seqreturn] ([yshift=\y cm]pl.south) --
    node[above, font=\scriptsize] {OpenQASM string}
    ([yshift=\y cm]comp.south);

  \pgfmathsetmacro{\y}{\ystart - 7.5*\ystep}
  \draw[seqarrow] ([yshift=\y cm]comp.south) -- ++(0.8, 0) -- ++(0, -0.5) -- ++(-0.8, 0)
    node[right, font=\scriptsize, xshift=0.9cm, yshift=0.25cm] {\texttt{\_remove\_gphase()}};

  \pgfmathsetmacro{\y}{\ystart - 8.5*\ystep}
  \draw[seqreturn] ([yshift=\y cm]comp.south) --
    node[above, font=\scriptsize] {QASM string (cleaned)}
    ([yshift=\y cm]user.south);

\end{tikzpicture}
\caption{Sequence diagram for \texttt{compile\_to\_qasm()}.  The loop
  iterates over every Blackbird operation, dispatching to the appropriate
  \texttt{CVOperation} subclass.}
\label{fig:sequence}
\end{figure}
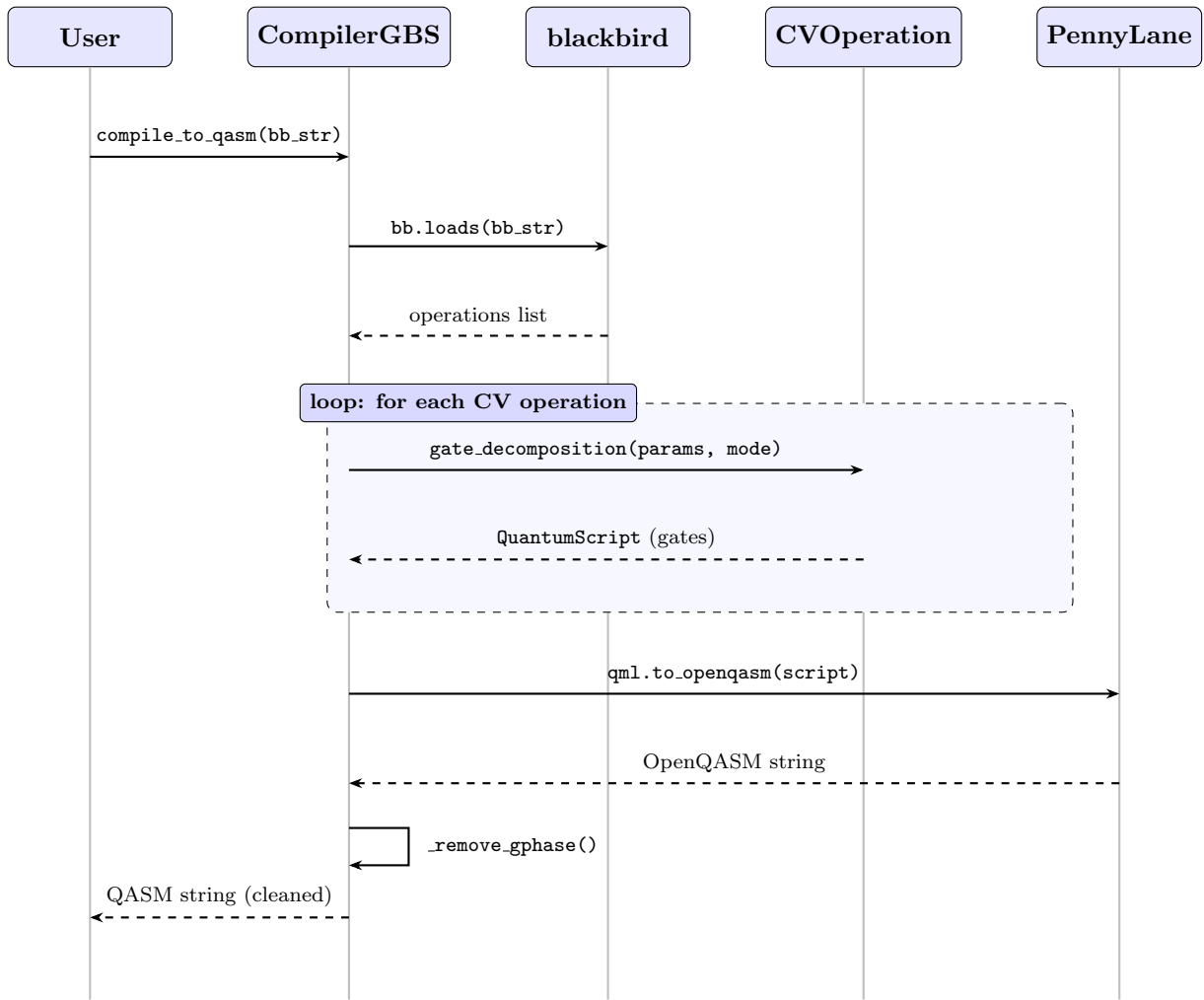

Fig.~\ref{fig:sequence} details the method-call sequence when the user
invokes \texttt{compile\_to\_qasm()}.

\subsection{Fock-Space Encoding}
\label{sec:encoding}

The mode-to-wire mapping is:
\begin{equation}
  \text{mode } m \mapsto
  \text{wires } \bigl[m \cdot Q,\; m \cdot Q + 1,\; \dots,\;
  (m+1)\cdot Q - 1\bigr],
\end{equation}
where $Q = \lceil\log_2 N\rceil$.

\subsection{Dispatch Mechanism}

\begin{figure}[h!]
\centering
\begin{tikzpicture}[node distance=0.4cm and 0.8cm]

  \node[draw, rounded corners, fill=orange!10, minimum width=2.5cm,
        minimum height=0.7cm, font=\small] (input) {Blackbird op dict};

  \node[draw, rounded corners, fill=gray!10, below=0.8cm of input,
        minimum width=6cm, minimum height=2.5cm] (table) {};
  \node[font=\small\bfseries, anchor=north] at ([yshift=-3pt]table.north)
    {\texttt{\_OP\_DISPATCH}};

  \node[font=\small\ttfamily, anchor=west] at ([xshift=5pt, yshift=0.4cm]table.west)
    {"Sgate" $\to$ \_compile\_sgate};
  \node[font=\small\ttfamily, anchor=west] at ([xshift=5pt, yshift=0.0cm]table.west)
    {"Dgate" $\to$ \_compile\_dgate};
  \node[font=\small\ttfamily, anchor=west] at ([xshift=5pt, yshift=-0.4cm]table.west)
    {"Rgate" $\to$ \_compile\_rgate};
  \node[font=\small\ttfamily, anchor=west] at ([xshift=5pt, yshift=-0.8cm]table.west)
    {"BSgate" $\to$ \_compile\_bsgate};

  \node[draw, rounded corners, fill=green!10, below=0.8cm of table,
        minimum width=2.5cm, minimum height=0.7cm, font=\small] (output)
    {QuantumScript (gates)};

  \draw[seqarrow] (input) -- node[right, font=\scriptsize] {lookup \texttt{op["op"]}} (table);
  \draw[seqarrow] (table) -- node[right, font=\scriptsize] {\texttt{getattr(self, method)(op)}} (output);

\end{tikzpicture}
\caption{Dictionary-based dispatch mechanism in \texttt{CompilerGBS}.}
\label{fig:dispatch}
\end{figure}
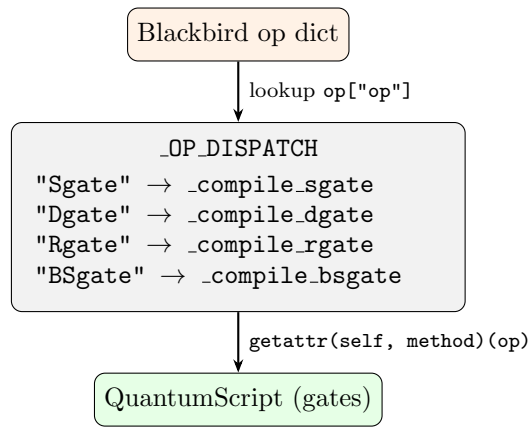

The compiler uses a dictionary-based dispatch to map Blackbird operation names
to compilation methods, Fig. \ref{fig:dispatch}. This avoids long \texttt{if/elif} chains and makes
adding new operations straightforward.

\subsection{Measurement Decoding}

After circuit execution, qubit measurement outcomes (binary strings) must be
converted back to photon-number samples.  The method
\texttt{binary\_to\_photon\_meas} performs this:

\begin{equation}
  n_m = \sum_{i=0}^{Q-1} q_{mQ+i} \cdot 2^{Q-1-i},
  \qquad m = 0, 1, \dots, M-1,
\end{equation}
where $M$ is the number of optical modes, $Q$ is the number of qubits per
mode, and $q_j \in \{0,1\}$ is the measurement outcome of qubit~$j$.

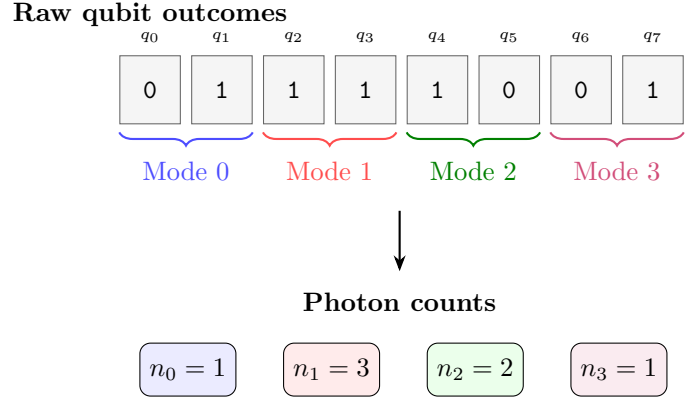
\begin{figure}[h!]
\centering
\begin{tikzpicture}[node distance=0.4cm]

  \node[font=\small\bfseries] at (0, 2.2) {Raw qubit outcomes};
  \foreach \i/\v in {0/0, 1/1, 2/1, 3/1, 4/1, 5/0, 6/0, 7/1} {
    \node[encbox, minimum width=0.8cm] (b\i) at (\i * 0.95, 1.2) {\v};
    \node[font=\tiny, above=1pt of b\i] {$q_{\i}$};
  }

  \draw[decorate, decoration={brace, amplitude=5pt, mirror}, thick, blue!70]
    ([yshift=-3pt]b0.south west) -- ([yshift=-3pt]b1.south east)
    node[midway, below=7pt, font=\small, blue!70] {Mode 0};
  \draw[decorate, decoration={brace, amplitude=5pt, mirror}, thick, red!70]
    ([yshift=-3pt]b2.south west) -- ([yshift=-3pt]b3.south east)
    node[midway, below=7pt, font=\small, red!70] {Mode 1};
  \draw[decorate, decoration={brace, amplitude=5pt, mirror}, thick, green!50!black]
    ([yshift=-3pt]b4.south west) -- ([yshift=-3pt]b5.south east)
    node[midway, below=7pt, font=\small, green!50!black] {Mode 2};
  \draw[decorate, decoration={brace, amplitude=5pt, mirror}, thick, purple!70]
    ([yshift=-3pt]b6.south west) -- ([yshift=-3pt]b7.south east)
    node[midway, below=7pt, font=\small, purple!70] {Mode 3};

  \draw[-{Stealth[length=6pt]}, thick] (3.3, -0.4) -- (3.3, -1.2);

  \node[font=\small\bfseries] at (3.3, -1.6) {Photon counts};
  \node[draw, fill=blue!8, rounded corners, minimum width=1.2cm, minimum height=0.7cm, font=\small]
    (p0) at (0.5, -2.5) {$n_0=1$};
  \node[draw, fill=red!8, rounded corners, minimum width=1.2cm, minimum height=0.7cm, font=\small]
    (p1) at (2.4, -2.5) {$n_1=3$};
  \node[draw, fill=green!8, rounded corners, minimum width=1.2cm, minimum height=0.7cm, font=\small]
    (p2) at (4.3, -2.5) {$n_2=2$};
  \node[draw, fill=purple!8, rounded corners, minimum width=1.2cm, minimum height=0.7cm, font=\small]
    (p3) at (6.2, -2.5) {$n_3=1$};

\end{tikzpicture}
\caption{Measurement decoding for a 4-mode circuit at cutoff~4 (8 qubits).
  Each pair of qubit outcomes is converted to a photon number via binary
  reconstruction.}
\label{fig:measurement}
\end{figure}

\subsection{Module Dependency Graph}

\begin{figure}[h!]
\centering
\hspace*{-1.5cm}\begin{tikzpicture}[
  pkg/.style={draw, rounded corners=4pt, fill=#1, minimum width=2.5cm,
              minimum height=0.9cm, font=\small\bfseries, align=center},
  pkg/.default={blue!10},
  node distance=0.3cm and 0.5,
]

  \node[pkg=blue!15] (init) {\texttt{gbs\_compiler}};
  \node[pkg=blue!10, below left=of init] (comp) {\texttt{compiler}};
  \node[pkg=blue!10, below right=of init] (ops) {\texttt{operations}};

  \draw[seqarrow] (init) -- (comp);
  \draw[seqarrow] (init) -- (ops);
  \draw[uses] (comp) -- (ops);

  \node[pkg=orange!15, below left=1.2cm and 0.3cm of comp] (bblib) {quantum-blackbird};
  \node[pkg=orange!15, below=1.2cm of comp] (pl) {PennyLane};
  \node[pkg=orange!15, below=1.2cm of ops] (sympy) {SymPy};
  \node[pkg=orange!15, below right=1.2cm and 0.3cm of ops] (numpy) {NumPy};

  \draw[uses] (comp) -- (bblib);
  \draw[uses] (comp) -- (pl);
  \draw[uses] (ops) -- (pl);
  \draw[uses] (ops) -- (sympy);
  \draw[uses] (ops) -- (numpy);

  \begin{scope}[on background layer]
    \node[draw=blue!40, dashed, rounded corners=8pt, fill=blue!3,
          fit=(init)(comp)(ops), inner sep=12pt, label={[font=\scriptsize\itshape]above:Internal}] {};
    \node[draw=orange!40, dashed, rounded corners=8pt, fill=orange!3,
          fit=(bblib)(pl)(sympy)(numpy), inner sep=12pt, label={[font=\scriptsize\itshape]below:External}] {};
  \end{scope}

\end{tikzpicture}
\caption{Module dependency graph.  Solid arrows = imports; dashed arrows = usage.}
\label{fig:deps}
\end{figure}
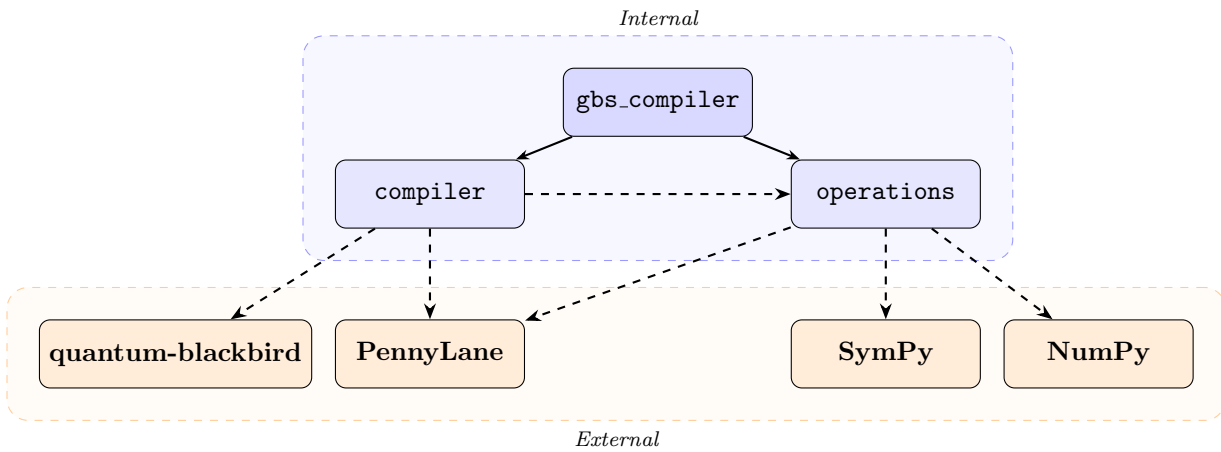

Fig.~\ref{fig:deps} shows the dependency relationships between the compiler
package and its external dependencies.

\subsection{Supported Operations}
Supported operations are summarized in Table \ref{tab:ops}.

\begin{table}[h!]
\centering
\begin{tabular}{llcccc}
  \toprule
  \textbf{Class} & \textbf{BB gate} & \textbf{Modes} &
    \textbf{Cutoff 2} & \textbf{Cutoff 4} & \textbf{Parameters} \\
  \midrule
  \texttt{PhaseShift}   & \texttt{Rgate}  & 1 & 1 $R_z$   & 2 $R_z$   & $\varphi$ \\
  \texttt{Displacement} & \texttt{Dgate}  & 1 & 3 gates   & $\sim$14 gates & $r,\,\varphi$ \\
  \texttt{Squeeze}      & \texttt{Sgate}  & 1 & Identity  & 6 gates   & $r,\,\varphi$ \\
  \texttt{BeamSplitter} & \texttt{BSgate} & 2 & 6 gates   & $\sim$58 gates & $\theta$, $\varphi$ \\
  \bottomrule
\end{tabular}
\caption{Operation summary.  Gate counts are approximate for cutoff~4 due to
  multi-controlled gate decomposition overhead.}
\label{tab:ops}
\end{table}

\section{Summary}

In this work, we have developed a systematic framework for representing 
fundamental quantum optical states and operations within a qubit-based 
quantum register. By introducing a controlled truncation of the optical 
Fock space, we established an explicit mapping between infinite-dimensional 
photonic Hilbert spaces and finite-dimensional qubit registers suitable for 
gate-based quantum computation.

We demonstrated that the truncated Fock space can be consistently realized using
appropriately defined creation and annihilation operators, preserving unitarity within the truncated Hilbert space and enabling a controlled representation of photonic dynamics within a well-defined low-photon-number regime. Furthermore, we showed that the finite-dimensional construction admits an algebraic interpretation in terms of \(q\)-deformed oscillator algebras at roots of unity, while the standard bosonic algebra is recovered in the infinite-dimensional limit.

Within this framework, we explicitly constructed qubit-level quantum circuits for
key optical operations, including phase rotations, displacement operations,
single-mode squeezing, and symmetric beam splitters. Detailed circuit
decompositions were presented for low-dimensional truncations, in particular
for $N=2$ and $N=4$, illustrating the feasibility of implementing nontrivial
quantum optical transformations on universal qubit-based quantum hardware.
The developed approach can be extended to higher truncation orders $N$, particularly 
$N=2^m$ with $m>2$. Such an extension, however, requires a new set of quantum circuits 
tailored to these truncations, building on the methodology established for the $N=4$ case.

The proposed approach enables direct simulation of truncated quantum optical circuits, 
including Gaussian Boson Sampling architectures, provided that the effects of the finite 
photon-number cut-off are taken into account. As such, it provides a concrete bridge 
between continuous-variable quantum optics and discrete-variable quantum computation, 
opening new possibilities for benchmarking photonic protocols and exploring hybrid 
quantum algorithms within finite-dimensional quantum registers.

A natural extension of the developed approach is the development of a 
reversed compilation process from the qubit-based to photonic circuits. 
We plan to address this in future studies. 

\section*{Acknowledgements}

The work has been performed within the ``Gaussian Boson Sampling 
for Earth Observation - GBS4EO'' project funded by the European Space 
Agency, under the contract No. 4000147776/25/I-DT-bgh. 
ESA classification: UNCLASSIFIED -- Releasable to the Public.
The project has been conducted in the consortium of KP Labs (leader),
Jagiellonian University and IT4Innovations (IT4I). In the research 
process, tools such as Mathematica, ChatGPT, and Claude have been used. 

 \subsection*{Appendix A: $q$-deformed oscillator}

Let us consider the following algebra of creation and 
annihilation operators:
\begin{align}
\hat{B}\hat{B}^\dagger - q \hat{B}^\dagger \hat{B} = \hat{\mathbb{I}}, 
\label{qdefalgebra}
\end{align}
where $q\in \mathbb{C}$ is a deformation parameter. Furthermore, 
let us introduce the occupation number operator $\hat{N}:=
\hat{B}^\dagger \hat{B}$, the eigenequation of which takes 
the following form:
\begin{equation}
\hat{N}_B \ket{n} = h(n) \ket{n},
\end{equation}
where, for now, $h(n)$ is some function of $n\in \mathbb{N}$ 
and $q$ to be determined. For this purpose, let us express the 
annihilation and creation operators as:
\begin{align}
\hat{B}\ket{n}=f(n)\ket{n-1}, \\
\hat{B}^{\dagger}\ket{n} = g(n) \ket{n+1},  
\end{align}
where $f(n)$ and $g(n)$ are some functions to be determined. 

From the definition of the occupation number operator, one reads that:
\begin{equation}
h(n) = f(n)g(n-1).
\end{equation}
Furthermore, from the $q$-deformed algebra (\ref{qdefalgebra}) 
one finds the relation:
\begin{equation}
f(n+1)g(n)-qf(n)g(n-1)=1,
\end{equation}
which can be rewritten into to the recurrence relation:
\begin{equation}
h(n+1)=qh(n)+1.
\end{equation}
The equation is simply a shifted relation for a geometric series.
By introducing the variable $S(n) = h(n)+\frac{1}{q-1}$, 
the equation rewrites to:
\begin{equation}
S(n+1)=qS(n),
\end{equation}
which is solved by $S(n)=q^nS(0).$ Returning to the variable $h(n)$ 
leads to:
\begin{equation}
h(n) = \frac{1-q^n}{1-q}+h(0)q^n,
\end{equation}
where $h(0)$ is fixed by the initial condition. Here, the condition 
corresponds to the undeformed limit ($q\rightarrow 1$), in which:
\begin{equation}
\lim_{n\rightarrow 1} h(n) = n+h(0),
\end{equation}
which fixes $h(0)=0$. Therefore, in consequence:
\begin{equation}
h(n) =  \frac{1-q^n}{1-q} := [n]_q.
\end{equation}
From here, one finds that:
\begin{align}
f(n) &= \sqrt{[n]_q},\\
g(n) &= \sqrt{[n+1]_q},
\end{align}
so that 
\begin{equation}
h(n) = f(n)g(n-1) = \sqrt{[n]_q}\sqrt{[n]_q} = [n]_q,
\end{equation}
and both $f(n)$ and $g(n)$ satisfy the correct 
undeformed limits ($q\rightarrow 1$), i.e.:
\begin{align}
\lim_{n\rightarrow 1} f(n) &= \sqrt{n},\\
\lim_{n\rightarrow 1} g(n) &= \sqrt{n+1}.
\end{align}

Furthermore, it is instructive to write the algebra of 
$\hat{B}, \hat{B}^{\dagger}$ and $\hat{N}_B$ using the 
standard commutators, which reads:
\begin{align}
[\hat{B},\hat{B}^{\dagger}] &=(q-1)\hat{N}_B+\hat{\mathbb{I}}, \\
[\hat{B},\hat{N}_B] &=  (q-1)\hat{N}_B\hat{B}+\hat{B},\\
[\hat{B}^{\dagger},\hat{N}_B] &= (q-1)\hat{B}^{\dagger}\hat{N}_B+\hat{B}^{\dagger},
\end{align}
so that in the $q\rightarrow 1$ limit the standard relations are recovered. 

\section*{Appendix B: Definitions of the quantum gates}

\begin{itemize}
\item $X, Y, Z$
\begin{equation}
X =
\begin{pmatrix}
0 & 1 \\
1 & 0
\end{pmatrix}, \ \ Y =
\begin{pmatrix}
0 & -i \\
i & 0
\end{pmatrix}, \ \ 
Z =
\begin{pmatrix}
1 & 0 \\
0 & -1
\end{pmatrix}
\end{equation}
\item $H$
\begin{equation}
    H=\frac{1}{\sqrt2}\begin{pmatrix}
        1&1\\
        1&-1\\
    \end{pmatrix}
\end{equation}
\item $S$
\begin{equation}
    S=R_z\left(\frac{\pi}{2}\right)=\begin{pmatrix}
        1&0\\
        0&i\\
    \end{pmatrix}
\end{equation}
\item $R_x$
\begin{equation}
R_x(\theta) =
\begin{pmatrix}
\cos\left(\frac{\theta}{2}\right) & -i\,\sin\left(\frac{\theta}{2}\right) \\
-i\,\sin\left(\frac{\theta}{2}\right) & \cos\left(\frac{\theta}{2}\right)
\end{pmatrix}
\end{equation}
\item $R_y$
\begin{equation}
R_y(\theta) =
\begin{pmatrix}
\cos\left(\frac{\theta}{2}\right) & -\sin\left(\frac{\theta}{2}\right) \\
\sin\left(\frac{\theta}{2}\right) & \cos\left(\frac{\theta}{2}\right)
\end{pmatrix}
\end{equation}
\item $R_z$
\begin{equation}
R_z(\theta) =
\begin{pmatrix}
e^{-i\theta/2} & 0 \\
0 & e^{i\theta/2}
\end{pmatrix}
\end{equation}
\item $U_1$
\begin{equation}U_1(\phi) = e^{i\phi/2}R_z(\phi) = \begin{pmatrix}
    1 & 0 \\
    0 & e^{i\phi}
\end{pmatrix}
\end{equation}
\item $U_2$
\begin{equation}U_2(\phi, \delta) = \frac{1}{\sqrt{2}}\begin{pmatrix} 1 & -\exp(i \delta)
\\ \exp(i \phi) & \exp(i (\phi + \delta)) \end{pmatrix}\end{equation}
\item $U_3$
\begin{equation}U_3(\theta, \phi, \delta) = \begin{pmatrix} \cos(\theta/2) & -\exp(i \delta)\sin(\theta/2) \\
\exp(i \phi)\sin(\theta/2) & \exp(i (\phi + \delta))\cos(\theta/2) \end{pmatrix}\end{equation}
\item $CNOT$ 
\begin{eqnarray}
 CNOT=\left(\begin{matrix}
 1 & 0 & 0 & 0 \\
 0 & 1 & 0 & 0 \\ 
 0 & 0 & 0 & 1 \\
 0 & 0 & 1 & 0
 \end{matrix}\right)
\end{eqnarray}
\item $SWAP$ 
\begin{eqnarray}
SWAP=\left(\begin{matrix}
 1 & 0 & 0 & 0 \\
 0 & 0 & 1 & 0 \\ 
 0 & 1 & 0 & 0 \\
 0 & 0 & 0 & 1
 \end{matrix}\right)
\end{eqnarray}
\item $G$ 
\begin{eqnarray}
 G(\theta)=\left(\begin{matrix}1 & 0 & 0 & 0\\0 & \cos{\left(\theta/2 \right)} & - \sin{\left(\theta/2 \right)} & 0\\0 & \sin{\left(\theta/2 \right)} & \cos{\left(\theta/2 \right)} & 0\\0 & 0 & 0 & 1\end{matrix}\right)
\end{eqnarray}

\end{itemize}

\bibliography{references}

\end{document}